\documentclass[pdflatex,sn-mathphys-ay]{sn-jnl}

\usepackage{graphicx}%
\usepackage{multirow}%
\usepackage{amsmath,amssymb,amsfonts}%
\usepackage{amsthm}%
\usepackage{mathrsfs}%
\usepackage[title]{appendix}%
\usepackage{xcolor}%
\usepackage{textcomp}%
\usepackage{manyfoot}%
\usepackage{booktabs}%
\usepackage{algorithm}%
\usepackage{algorithmicx}%
\usepackage{algpseudocode}%
\usepackage{listings}%

\usepackage[english]{babel}
\usepackage{bbm}
\usepackage{mathtools}
\usepackage{braket}
\usepackage{color}
\usepackage{comment}
\usepackage{dsfont}
\usepackage{lettrine}
\usepackage{units}
\usepackage{array}
\usepackage{hhline}

\begin{document}

\title[Quantum Neural Networks for Cloud Cover Parameterizations in Climate Models]{Quantum Neural Networks for Cloud Cover Parameterizations in Climate Models}

\author*[1]{\fnm{Lorenzo} \sur{Pastori}}\email{lorenzo.pastori@dlr.de}

\author[1]{\fnm{Arthur} \sur{Grundner}}\email{arthur.grundner@dlr.de}

\author[1,2]{\fnm{Veronika} \sur{Eyring}}\email{veronika.eyring@dlr.de}

\author[1]{\fnm{Mierk} \sur{Schwabe}}\email{mierk.schwabe@dlr.de}

\affil[1]{\orgdiv{Institut f\"ur Physik der Atmosph\"are}, \orgname{Deutsches Zentrum für Luft- und Raumfahrt (DLR)}, \orgaddress{\city{Oberpfaffenhofen}, \country{Germany}}}

\affil[2]{\orgdiv{Institute of Environmental Physics (IUP)}, \orgname{University of Bremen}, \orgaddress{\city{Bremen}, \country{Germany}}}

\abstract{Long-term climate projections require running global Earth system models on timescales of hundreds of years and have relatively coarse resolution (from 40 to 160 km in the horizontal) due to their high computational costs. Unresolved subgrid-scale processes, such as clouds, are described in a semi-empirical manner by so called parameterizations, which are a major source of uncertainty in climate projections. Machine learning models trained on short high-resolution climate simulations are  promising candidates to replace conventional parameterizations. In this work, we take a step further and explore the potential of quantum machine learning, and in particular quantum neural networks (QNNs), to develop cloud cover parameterizations. QNNs differ from their classical counterparts, and their potentially high expressivity turns them into promising tools for accurate data-driven schemes to be used in climate models. Here we perform an extensive comparative analysis between several QNNs and classical neural networks (NNs), by training both ansatzes on data coming from high-resolution simulations with the ICOsahedral Non-hydrostatic weather and climate model (ICON). Our results show that the overall performance of the investigated QNNs is comparable to that of classical NNs of similar size, i.e., with the same number of trainable parameters, with both ansatzes outperforming standard parameterizations used in climate models. Our study also includes an analysis of the generalization ability of the models as well as the geometrical properties of their optimization landscape. We furthermore investigate the effects of finite sampling noise, and show that the training and the predictions of the QNNs are stable even in this noisy setting. These results demonstrate the applicability of quantum machine learning models to learn meaningful patterns in climate data, and are thus relevant for a broad range of problems within the climate modeling community.}

\keywords{Quantum machine learning, Quantum neural networks, Climate modeling, Parameterizations, Cloud cover}

\maketitle

\section{Introduction} \label{sec:introduction}
One of the requirements for improving mitigation and adaption strategies against climate change is the ability to perform reliable long-term climate projections using climate models. Climate models are numerical models that simulate the evolution of the various components of the Earth system \citep{Jacobson2005,Gettelman2016}. In a climate model, the equations governing the dynamics of the atmosphere are discretized on a grid with horizontal extension on the order of tens of kilometers, to enable ensembles of climate projections over several decades. Due to this relatively coarse horizontal resolution, current climate models are still affected by systematic biases compared to observations, despite continuous improvements \citep{Sherwood2014,Bock2020}. At these horizontal resolutions, important physical processes such as clouds, convection or turbulence cannot be resolved, and their effect must be re-introduced in the climate model in an approximate manner by so-called parameterizations \citep{Stensrud2007,McFarlane2011,Christensen2022}. The structural and parametric uncertainties in conventional parameterization schemes are a source of the aforementioned remaining biases \citep{Randall2003,Sherwood2014,Schneider2017,Eyring2021a}. Machine learning (ML) models are promising candidates to replace conventional parameterizations \citep{Gentine2021,Eyring2021a,Eyring2024a,Eyring2024b}. Several uses of ML for parameterizations have been proposed in the literature, with prominent applications to radiation \citep{Chevallier1998,Krasnopolsky2005,Lagerquist2021,Hafner2024}, convection \citep{Krasnopolsky2013,Rasp2018,Gentine2018,Brenowitz2019,Yuval2020,Heuer2024} and cloud cover \citep{Grundner2022,Grundner2024}.  One of the strategies in employing ML is to develop data-driven schemes trained on coarse-grained data coming from high-resolution climate simulations, where convection and clouds are more explicitly resolved. While enabling numerous improvements, the use of ML also introduces several requirements, such as the need of complex yet trainable models to encompass various physical scenarios, the need of large amounts of training data, and the ability to generalize to unseen climate regimes \citep{Eyring2024a}.

In this work, we explore the potential of quantum machine learning (QML) \citep{PerdomoOrtiz2018,Benedetti2019,Cerezo2022}, and in particular quantum neural networks (QNNs) \citep{Farhi2018}, to develop parameterizations for climate models, specifically focusing on the case of cloud cover. QNNs constitute a different type of ansatz compared to classical neural networks (NNs), and several theoretical works have highlighted their good generalization capability \citep{Caro2021,Banchi2021,Caro2022a,Haug2024}, their higher expressivity for certain tasks \citep{Du2020,Yu2023}, as well as provided hints to their well-behaved optimization landscape which may result in a good trainability in certain regimes \citep{Abbas2021}. Motivated by these insights, we perform a thorough comparative analysis between several QNNs and classical NNs models, to understand whether this type of QML models can help addressing the aforementioned challenges.

The field of QML expands in several directions, including supervised learning for regression and classification \citep{Farhi2018}, generative modeling \citep{Amin2018,DallaireDemers2018,Coyle2020,Gao2022} and kernel methods \citep{Havlicek2019,Schuld2019a}. The number of QML applications to classical problems and datasets is rapidly growing (see \cite{Chen2022,Hur2022,Shen2024,Belis2024,Corli2024,Duneau2024,Aizpurua2024,Suenkel2023,Sakhnenko2024,Slabbert2024}, to name a few), thus potentially increasing its scope beyond purely quantum-related problems. Not surprisingly, there is also growing interest in the application of quantum computing and QML in the context of weather and climate science \citep{Tennie2023,Lachure2023,Nivelkar2023,Otgonbaatar2023,Nammouchi2023,Rahman2024,Jaderberg2024,Matsuta2024,Ho2024,Bazgir2024,Schwabe2024}. The application of QML to classical data is still in its infancy, also due to the noise and limited size of current noisy intermediate-scale quantum (NISQ) devices. The tests reported in the literature provide an empirical standpoint of the field, and it is still unclear how much quantum advantage can be distilled from QML in the long run \citep{Koelle2024,Cerezo2024,Bowles2024,Gilfuster2024,Bermejo2024}. Nevertheless, it is worth to investigate whether QML models lend themselves to learning patterns in climate-specific data, and whether they can do so better than classical architectures, for potentially bringing quantum or quantum-inspired enhancements into climate projections.

Here, we conduct such an investigation in the context of climate model parameterizations. Specifically, we perform numerical simulations training and evaluating both QNN and classical NN ansatzes on coarse-grained data coming from high-resolution simulations with the ICOsahedral Non-hydrostatic weather and climate model (ICON) \citep{Giorgetta2018}, which were part of the DYnamics of the Atmospheric general circulation Modeled On Non‐hydrostatic Domains (DYAMOND) project \citep{Stevens2019,Duras2021}. Both ansatzes are constructed to take as input the state variables in the climate model’s cell, such as humidity, temperature and pressure, and to output the cloud cover, i.e., the fraction of the cell that is occupied by clouds. The models developed here are trained and evaluated offline, i.e., without coupling them with the dynamical core of the climate model. These quantum and classical models are compared on several aspects. First, we focus on the prediction accuracy of the two types of ansatzes, and show that both achieve comparable accuracy when the number of trainable parameters is kept the same, while outperforming the conventional cloud cover scheme used in ICON. We then compare the generalization capabilities of the two ansatzes, in terms of the number of training data required to achieve a certain prediction accuracy, and the trainability, i.e., the number of training epochs required for the network (classical or quantum) to reach its optimal configuration. The average prediction performance of quantum and classical models is comparable also w.r.t.~these two aspects, despite quantum models having better behaved geometrical properties captured by the Fisher information matrix. Finally, in view of a practical implementation on quantum devices, we investigate the effects of finite sampling noise on the training and the predictions of QNNs. We show that our QNNs can train stably even in presence of shot noise, and that the costs of training and inference in terms of number of measurement shots can be further minimized using recently introduced variance regularization techniques \citep{Kreplin2024}. Our study thus highlights an important and timely use case of QML while also discussing its potential limitations.

The paper is organized as follows. In Section \ref{sec:training_data} we describe the training data used. In Section \ref{sec:QNN_parameterizations} we describe our design choices for the QNNs used, and provide details on our data encoding strategy as well as on the training of these models. In Section \ref{sec:results_noiseless} we present our results in the noiseless regime, focusing on the aspects of prediction accuracy, generalization and trainability, and including the comparisons with classical NNs. In Section \ref{sec:results_noisy} we present the analysis of our QNNs in presence of finite sampling noise. We finally conclude in Section \ref{sec:discussion}.

\section{The training data} \label{sec:training_data}

\begin{figure*}
    \centering
    \includegraphics[width=\linewidth]{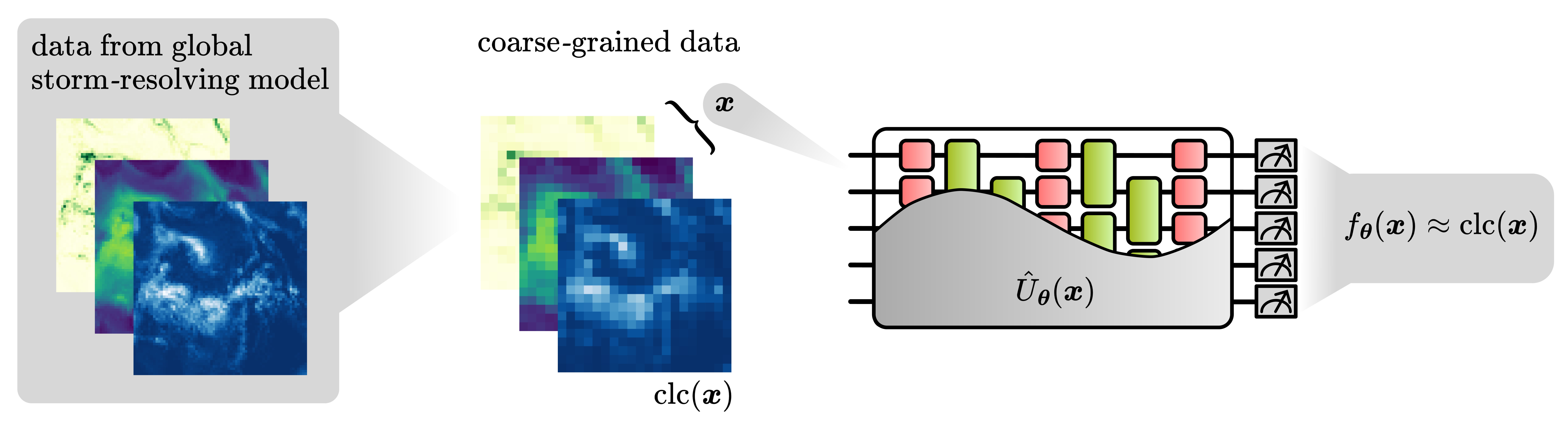}
    \caption{Schematics of our approach for developing a QNN-based parameterization. We coarse-grain high-resolution data to the target resolution, and construct a training dataset with coarse-grained state variables $\boldsymbol{x}$ as inputs and the corresponding coarse-grained cloud cover $\mathrm{clc}(\boldsymbol{x})$ as output. The dataset is used to train a QNN, i.e., to optimize the parameters $\boldsymbol{\theta}$ so that the QNN output $f_{\boldsymbol{\theta}}(\boldsymbol{x})$ approximates $\mathrm{clc}(\boldsymbol{x})$.}
    \label{fig:QNN_param_schematics} 
\end{figure*}

The training data used in this work is obtained from global high-resolution ICON simulations, from the DYAMOND project \citep{Stevens2019,Duras2021,Stephan2022}. These simulations offer an improved representation of clouds and convection compared to simulations at climate model resolutions \citep{Stevens2020}. They consist of 40 simulated days starting on the 1st August 2016, and 40 simulated days starting on the 20th January 2020, with a resolution of approximately 2.5 km in the horizontal. In both cases, the first 10 days have been discarded as spin-up time of the simulation, to have training and testing datasets more closely representing physically realistic conditions.

Following \cite{Giorgetta2022} we define a high-resolution grid cell to be cloudy (cloud cover $=1$) whenever a meaningful cloud condensate (cloud water or cloud ice) amount is detected (i.e., when specific cloud condensate content exceeds $10^{-6}$ kg/kg) and to otherwise be cloud-free (cloud cover $=0$). Such a binary setting of cloud cover is much more sensible at the high horizontal and vertical resolution of the storm-resolving model simulations than at coarse climate model resolutions.

The data is then coarse-grained to a horizontal resolution of approximately 80 km (corresponding to an R2B5 ICON grid), and vertically from 58 to 27 model levels below an altitude of 21 km, following \cite{Grundner2022,Grundner2024}. After coarse-graining, cloud cover in a given cell can take any value between 0 and 1, representing the fraction of the cell that is occupied by clouds. Given that cloud cover cannot exist in the absence of cloud condensate, we remove from the dataset all the cells where the total amount of cloud condensate is zero. This results in a dataset which is more balanced, i.e., where the cloud-free samples are less over-represented.

\section{Quantum neural networks as parameterization models} \label{sec:QNN_parameterizations}

\begin{figure*}
    \centering
    \includegraphics[width=\linewidth]{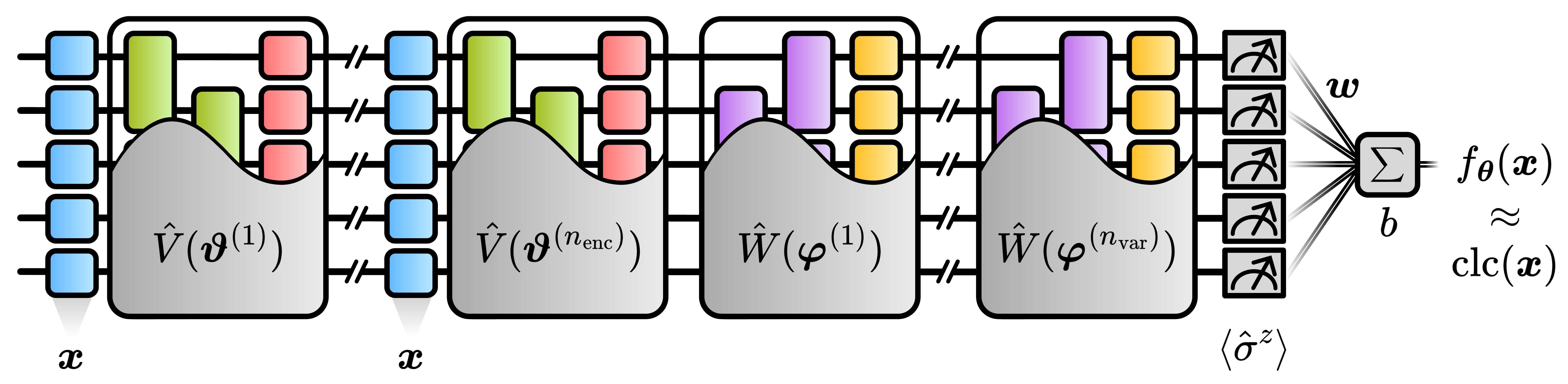}
    \caption{Schematics of our QNN architecture. The data $\boldsymbol{x}$ is uploaded $n_{\mathrm{enc}}$ times as angles of single-qubit rotations (blue boxes). In our implementation, each input feature is uploaded to the same qubit each time. These re-uploading gates are interleaved with variational blocks $\hat{V}(\boldsymbol{\vartheta}^{(k)})$ containing entangling gates and trainable parameters $\boldsymbol{\vartheta}^{(k)}$ ($k=1,\dots,n_{\mathrm{enc}}$). Afterwards, a sequence of $n_{\mathrm{var}}$ variational blocks $\hat{W}(\boldsymbol{\varphi}^{(\ell)})$ ($\ell=1,\dots,n_{\mathrm{var}}$) are applied. In the end, the expectation values of $\hat{\sigma}^z$ on all qubits are measured, and a weighted average of those is performed, with trainable weights $\boldsymbol{w}$ and a bias term $b$. The result $f_{\boldsymbol{\theta}}(\boldsymbol{x})$ should approximate $\mathrm{clc}(\boldsymbol{x})$ after training the parameters $\boldsymbol{\theta}=\{\{\boldsymbol{\vartheta}^{(k)}\}_k, \{\boldsymbol{\varphi}^{(\ell)}\}_{\ell}, \boldsymbol{w}, b\}$.}
    \label{fig:QNN_arch_schematics} 
\end{figure*}

The approach we use to construct our QNN-based parameterization for cloud cover is summarized in Fig.~\ref{fig:QNN_param_schematics}. The data resulting from the coarse-graining procedure described above constitute the training dataset $\mathcal{D}=\{\boldsymbol{x}_i,y_i\}_{i}$, where $\boldsymbol{x}_i\in\mathbb{R}^N$ denotes the selected coarse-grained state variables used as inputs and $y_i\in\mathbb{R}$ the output corresponding to the coarse-grained cloud cover $\mathrm{clc}(\boldsymbol{x}_i)$, for the $i$-th model cell (at a specific spatial location and point in time). These data are used for training our QNNs, which are based on a parameterized quantum circuit realizing the unitary operator $\hat{U}_{\boldsymbol{\theta}}(\boldsymbol{x})$, depending on the inputs $\boldsymbol{x}$ and on the union of all sets of trainable parameters $\boldsymbol{\theta}\in\mathbb{R}^D$. During optimization, all parameters in $\boldsymbol{\theta}$ are adjusted so that the distance between the QNN output $f_{\boldsymbol{\theta}}(\boldsymbol{x})$ and $\mathrm{clc}(\boldsymbol{x})$ for all $\boldsymbol{x}\in\{\boldsymbol{x}_i\}_i$ is minimized. In the following, we discuss the details of our implementations of $\hat{U}_{\boldsymbol{\theta}}(\boldsymbol{x})$ and how the QNN output $f_{\boldsymbol{\theta}}(\boldsymbol{x})$ is constructed.

\subsection{Architecture choice} \label{subsec:QNN_architecture}
A schematic visualization of the parameterized quantum circuit (PQC) forming the backbone of our QNNs is shown in Fig.~\ref{fig:QNN_arch_schematics}. The qubit register is initialized in the $|0\rangle$ state. To encode the input features (i.e., the components of the vector $\boldsymbol{x}$), we adopt the data re-uploading technique \citep{PerezSalinas2020,Schuld2021}, to increase the number of Fourier frequencies that our model can capture \citep{Schuld2021,Casas2023}. Specifically, we encode the input features multiple times as angles of single-qubit rotations, which results in the encoding layer
\begin{equation}
    \hat{S}(\boldsymbol{x})=\prod_{n=1}^N\mathrm{e}^{-\mathrm{i}\frac{x_n}{2}\hat{\sigma}^{\alpha}_n}\equiv\hat{R}_{\alpha}(\boldsymbol{x}) \,\,,
\end{equation}
where $x_n$ is the $n$-th component of $\boldsymbol{x}$ and $\alpha=x,y,z$ denotes the rotation axis, depending on the specific ansatz. In this construction, the number of qubits is therefore equal to the number of input features $N$. Subsequent applications of $\hat{S}(\boldsymbol{x})$ are interleaved with variational blocks $\hat{V}(\boldsymbol{\vartheta}^{(k)})$ (with $k=1,...,n_{\mathrm{enc}}$), which depend on trainable parameters $\boldsymbol{\vartheta}^{(k)}$ and also contain entangling operations. The specific form of these blocks is specified later on in this section. After this stage, amounting to $n_{\mathrm{enc}}$ data re-uploads, additional $n_{\mathrm{var}}$ variational blocks $\hat{W}(\boldsymbol{\varphi}^{(\ell)})$ (with $\ell=1,...,n_{\mathrm{var}}$) are applied to increase the number of trainable parameters. These blocks depend on trainable parameters $\boldsymbol{\varphi}^{(\ell)}$ and also contain entangling operations. The resulting unitary operator describing the PQC reads as
\begin{equation}
    \hat{U}_{\boldsymbol{\vartheta},\boldsymbol{\varphi}}(\boldsymbol{x})=\prod_{\ell=1}^{n_{\mathrm{var}}}\hat{W}(\boldsymbol{\varphi}^{(\ell)}) \prod_{k=1}^{n_{\mathrm{enc}}}\Big(\hat{V}(\boldsymbol{\vartheta}^{(k)})\hat{S}(\boldsymbol{x})\Big) \,\,,
\end{equation}
where the subscripts $\boldsymbol{\vartheta},\boldsymbol{\varphi}$ denote the dependences on the sets $\boldsymbol{\vartheta}=\{\boldsymbol{\vartheta}^{(k)}\}_k$ and $\boldsymbol{\varphi}=\{\boldsymbol{\varphi}^{(\ell)}\}_{\ell}$. After the PQC computation, the expectation values 
\begin{equation*}
\langle\hat{\sigma}_n^z\rangle_{\boldsymbol{\vartheta},\boldsymbol{\varphi}}(\boldsymbol{x})=\langle 0|\hat{U}^{\dagger}_{\boldsymbol{\vartheta},\boldsymbol{\varphi}}(\boldsymbol{x})\,\hat{\sigma}_n^z\,\hat{U}_{\boldsymbol{\vartheta},\boldsymbol{\varphi}}(\boldsymbol{x})|0\rangle 
\end{equation*}
are measured on the output state, and their weighted average with trainable weights $\boldsymbol{w}$ is computed, finally yielding the QNN prediction as
\begin{equation}
    f_{\boldsymbol{\theta}}(\boldsymbol{x})=b+\sum_{n=1}^N w_n\,\langle\hat{\sigma}_n^z\rangle_{\boldsymbol{\vartheta},\boldsymbol{\varphi}}(\boldsymbol{x}) \,\,,
    \label{eq:final_QNN_pred}
\end{equation}
with $w_n$ being the $n$-th component of $\boldsymbol{w}$, $b$ an additional trainable parameter (bias), and $\boldsymbol{\theta}$ summarizing all the parameters $\boldsymbol{\vartheta}$, $\boldsymbol{\varphi}$, $\boldsymbol{w}$, and $b$.

In this work, we focus on two ansatzes for the encoding and variational blocks in $\hat{U}_{\boldsymbol{\vartheta},\boldsymbol{\varphi}}(\boldsymbol{x})$. We label the resulting QNN architectures with XYZ and ZZXY. In both architectures, the encoding layer is implemented as $\hat{S}(\boldsymbol{x})=\hat{R}_{x}(\boldsymbol{x})$. For the XYZ circuit ansatz, the encoding blocks take the following form
\begin{equation*}
\begin{split}
    \hat{V}_{\mathrm{XYZ}}(\boldsymbol{\vartheta})=\,&\hat{R}_{yy}(\boldsymbol{\vartheta}_{(2N-1)\to(3N-3)})\,\hat{R}_{xx}(\boldsymbol{\vartheta}_{N\to(2N-2)})\times \\
    &\hat{R}_{zz}(\boldsymbol{\vartheta}_{1\to(N-1)}) \,\,,
\end{split}
\end{equation*}
where $\hat{R}_{\alpha\alpha}(\boldsymbol{\vartheta})=\prod_{n=1}^{N-1}\mathrm{e}^{-\mathrm{i}\frac{\vartheta_n}{2}\hat{\sigma}^{\alpha}_n\hat{\sigma}^{\alpha}_{n+1}}$, with $\alpha=x,y,z$, and $\boldsymbol{\vartheta}_{i\to j}$ denoting the slice of $\boldsymbol{\vartheta}$ from the $i$-th to $j$-th component. The variational blocks for XYZ read as
\begin{equation*}
\begin{split}
    \hat{W}_{\mathrm{XYZ}}(\boldsymbol{\varphi})=\,&\hat{R}_{x}(\boldsymbol{\varphi}_{(3N-2)\to(4N-3)})\times \\
    &\hat{R}_{yy}(\boldsymbol{\varphi}_{(2N-1)\to(3N-3)})\times \\
    &\hat{R}_{xx}(\boldsymbol{\varphi}_{N\to(2N-2)})\,\hat{R}_{zz}(\boldsymbol{\varphi}_{1\to(N-1)}) \,\,.
\end{split}
\end{equation*}
For the ZZXY ansatz the encoding and variational blocks take the following form
\begin{equation*}
    \hat{V}_{\mathrm{ZZXY}}(\boldsymbol{\vartheta})=\hat{R}_{y}(\boldsymbol{\vartheta}_{N\to(2N-1)})\,\hat{R}_{zz}(\boldsymbol{\vartheta}_{1\to(N-1)}) \,\,,
\end{equation*}
and
\begin{equation*}
\begin{split}
    \hat{W}_{\mathrm{ZZXY}}(\boldsymbol{\varphi})=\,&\hat{R}_{y}(\boldsymbol{\varphi}_{2N\to(3N-1)})\,\hat{R}_{zz}(\boldsymbol{\varphi}_{(N+1)\to(2N-1)})\times \\
    &\hat{R}_{x}(\boldsymbol{\varphi}_{1\to N}) \,\,.
\end{split}
\end{equation*}
We refer the reader to Appendix \ref{app:other_QNNs} for a comparison with other investigated types of QNNs. In this work, the investigated QNNs are numerically simulated in Python using the Pennylane library \citep{Bergholm2022} and optimized with JAX \citep{jax}.

\subsection{Input features and pre-processing} \label{subsec:QNN_inputs}
The components of the input vector $\boldsymbol{x}$ are the coarse-grained state variables chosen as predictors for the cloud cover in the corresponding model cell. These are chosen among the following state variables
\begin{itemize}
    \item $q_{\mathrm{v}}$ [kg/kg]: specific humidity,
    \item $q_{\mathrm{c}}$ [kg/kg]: specific cloud water content,
    \item $q_{\mathrm{i}}$ [kg/kg]: specific cloud ice content,
    \item $T$ [K]: air temperature,
    \item $p$ [Pa]: pressure,
    \item $h_{\mathrm{w}}=\sqrt{u^2+v^2}$ [m/s]: magnitude of horizontal wind component (with $u$ and $v$ being the zonal and meridional components, respectively),  
    \item $z_{\mathrm{g}}$ [m]: geometric height at full level,
    \item $\phi$ [rad]: latitude.
\end{itemize}
The selection of these variables is based on previous works on ML-based cloud cover parameterizations \citep{Grundner2022,Grundner2024}. Given the different magnitudes and distributions of these input features, it is necessary to suitably transform and re-scale them so that they can be encoded as angles in our PQCs. For the features $T$, $p$ and $z_{\mathrm{g}}$ we found it sufficient to perform a min-max (linear) re-scaling within the interval $[0,\pi]$. For the features $q_{\mathrm{v}}$, $q_{\mathrm{c}}$, $q_{\mathrm{i}}$ and $h_{\mathrm{w}}$ the situation is slightly more complex, since their distribution is sharply peaked at $0$ and contains long decaying tails, in particular for $q_{\mathrm{c}}$, $q_{\mathrm{i}}$. For these features we perform a non-linear transformation specifically constructed to (i) make the input feature distribution more uniform, so as to better distinguish from each other the values close to $0$, and (ii) retain the input feature variability in the tails, which can be associated with physical scenarios we are interested in capturing. In Appendix \ref{app:input_transf} we provide the details on the transformations we constructed keeping these objectives in mind. The transformed features $q_{\mathrm{v}}$, $q_{\mathrm{c}}$, $q_{\mathrm{i}}$ and $h_{\mathrm{w}}$ are also approximately distributed in the interval $[0,\pi]$, so that they can be used as angles in the PQCs. To enable a proper comparison between quantum and classical NNs, the same input transformations are used in both cases.

As an additional pre-processing step further enhancing the performance of our networks, it is beneficial to learn a transformed version of cloud cover, i.e., to set the outputs $y_i$ in the training set to $y_i=g(\mathrm{clc}(\boldsymbol{x}_i))$, with $g$ denoting a suitable transformation function. The transformation function $g$ is constructed to have the training outputs $y_i$ approximately uniformly distributed in the interval $[0,1]$. An explicit expression of this transformation is given in Appendix \ref{app:input_transf}.

\subsection{Training QNNs} \label{subsec:QNN_training}
In an actual implementation of QNNs on quantum devices, the training of the parameters $\boldsymbol{\theta}$ is achieved via a quantum-classical feedback loop \citep{McClean2016,Cerezo2021}. At each iteration of this loop the QNN is run on the quantum device for given parameters $\boldsymbol{\theta}$ and the cost function is computed, and its value is used to propose new parameters to be used in the QNN at the next iteration. In our case, the computations of the QNN which would take place on a quantum device are simulated numerically. The cost function we minimize for training is the mean squared error (MSE) over the training dataset $\mathcal{D}$ of size $N_{\mathrm{train}}=|\mathcal{D}|$, computed as
\begin{equation}
    \mathrm{MSE}_{\mathcal{D}}(\boldsymbol{\theta})=\frac{1}{N_{\mathrm{train}}}\sum_{i=1}^{N_{\mathrm{train}}}\Big(f_{\boldsymbol{\theta}}(\boldsymbol{x}_i)-y_i)\Big)^2 \,\,,
    \label{eq:MSEdef}
\end{equation}
with $y_i=g(\mathrm{clc}(\boldsymbol{x}_i))$. The parameters $\boldsymbol{\theta}$ are updated using gradient descent methods. Specifically, we use the Adam optimizer \citep{Kingma2014}. This requires the ability of calculating the gradients of the cost function with respect to the parameters efficiently, which for QNNs can be done using the parameter-shift rule \citep{Mitarai2018,Schuld2019}. In our case all the gate generators are (one half times) Pauli operators multiplied by the variational parameters, and we can use the following form for the parameter-shift rule
\begin{equation}
    \frac{\partial\langle\hat{O}\rangle_{\boldsymbol{\vartheta}}}{\partial\phi_j}=\frac{1}{2}\Big(\langle\hat{O}\rangle_{\boldsymbol{\vartheta}+\frac{\pi}{2}\boldsymbol{e}_j}-\langle\hat{O}\rangle_{\boldsymbol{\vartheta}-\frac{\pi}{2}\boldsymbol{e}_j}\Big) \,\,,
    \label{eq:param_shift_rule}
\end{equation}
with $\langle\hat{O}\rangle_{\boldsymbol{\vartheta}}=\langle 0|\hat{U}^{\dagger}_{\boldsymbol{\vartheta}}\,\hat{O}\,\hat{U}_{\boldsymbol{\vartheta}}|0\rangle$, where $\hat{O}$ is a generic observable and $\hat{U}_{\boldsymbol{\vartheta}}$ is the unitary corresponding to the PQC with parameters $\boldsymbol{\vartheta}$ (analogous rules for the derivatives w.r.t.~$\boldsymbol{\varphi}$ apply), where we dropped the possible dependence on the inputs $\boldsymbol{x}$. Applying this formula to our instances of PQCs, we obtain the gradients of the expectation values $\langle\hat{\sigma}_n^z\rangle_{\boldsymbol{\vartheta},\boldsymbol{\varphi}}(\boldsymbol{x})$ and use them for computing the derivatives of Eq. (\ref{eq:MSEdef}).

\section{Results in the noiseless regime} \label{sec:results_noiseless}

\begin{table}
\centering
\begin{tabular}{|l|l|l|l|l|l|}
\hline 
QNN & $N$ & $n_{\mathrm{enc}}$ & $n_{\mathrm{var}}$ & $D$ & Input features \\ \hline\hline
$\mathrm{XXY}^8_{5,3}$ & 8 & 5 & 3 & 201 & $\{q_{\mathrm{v}}, q_{\mathrm{c}}, q_{\mathrm{i}}, T, p, z_{\mathrm{g}}, h_{\mathrm{w}}, \phi\}$ \\ 
\hline
$\mathrm{ZZXY}^8_{2,7}$ & 8 & 2 & 7 & 200 & $\{q_{\mathrm{v}}, q_{\mathrm{c}}, q_{\mathrm{i}}, T, p, z_{\mathrm{g}}, h_{\mathrm{w}}, \phi\}$ \\ 
\hline
$\mathrm{XXY}^6_{4,2}$ & 6 & 4 & 2 & 109 & $\{q_{\mathrm{v}}, q_{\mathrm{c}}, q_{\mathrm{i}}, T, p, h_{\mathrm{w}}\}$ \\ 
\hline
$\mathrm{ZZXY}^6_{2,5}$ & 6 & 2 & 5 & 114 & $\{q_{\mathrm{v}}, q_{\mathrm{c}}, q_{\mathrm{i}}, T, p, h_{\mathrm{w}}\}$ \\ 
\hline
\end{tabular}

\vspace{5pt}

\begin{tabular}{|l|l|l|l|l|l|}
\hline 
NN & Hidden layers & $D$ & Input features \\ \hline\hline
$\mathrm{NN}^8_{12,6,2}$ & $12\to6\to2$ & 203 & $\{q_{\mathrm{v}}, q_{\mathrm{c}}, q_{\mathrm{i}}, T, p, z_{\mathrm{g}}, h_{\mathrm{w}}, \phi\}$ \\ 
\hline
$\mathrm{NN}^6_{8,3,7}$ & $8\to3\to7$ & 119 & $\{q_{\mathrm{v}}, q_{\mathrm{c}}, q_{\mathrm{i}}, T, p, h_{\mathrm{w}}\}$ \\ 
\hline
\end{tabular}
\vspace{5pt}
\caption{Summary of the quantum (upper table) and classical (lower table) architectures used in the main text. For the quantum models, $N$ refers to the number of qubits, corresponding to the number of features, and $n_{\mathrm{enc}}$ and $n_{\mathrm{var}}$ to the number of encoding and variational blocks in the PQC, respectively. For the classical models, the number of nodes in the hidden NN layers is shown (the first `visible' layer is the input layer with $N$ input nodes, and the last layer is the output layer with one node). Both classical NNs use tanh activations. $D$ is the number of trainable parameters.}
\label{table:QNNarchs}
\end{table}

In this section we present the results from our QNNs in the absence of sampling noise, i.e., with the expectation values $\langle\hat{\sigma}_n^z\rangle_{\boldsymbol{\vartheta},\boldsymbol{\varphi}}(\boldsymbol{x})$ evaluated to numerical precision. This allows us to more directly assess the learning and representational capabilities of our networks in the context of cloud cover parameterizations, and to make a better comparison with other existing classical schemes.

\subsection{Evaluation of QNN predictions and comparison with classical schemes} \label{subsec:noiseless_training_and_eval}

\begin{figure*}
    \centering
    \includegraphics[width=1.05\linewidth]{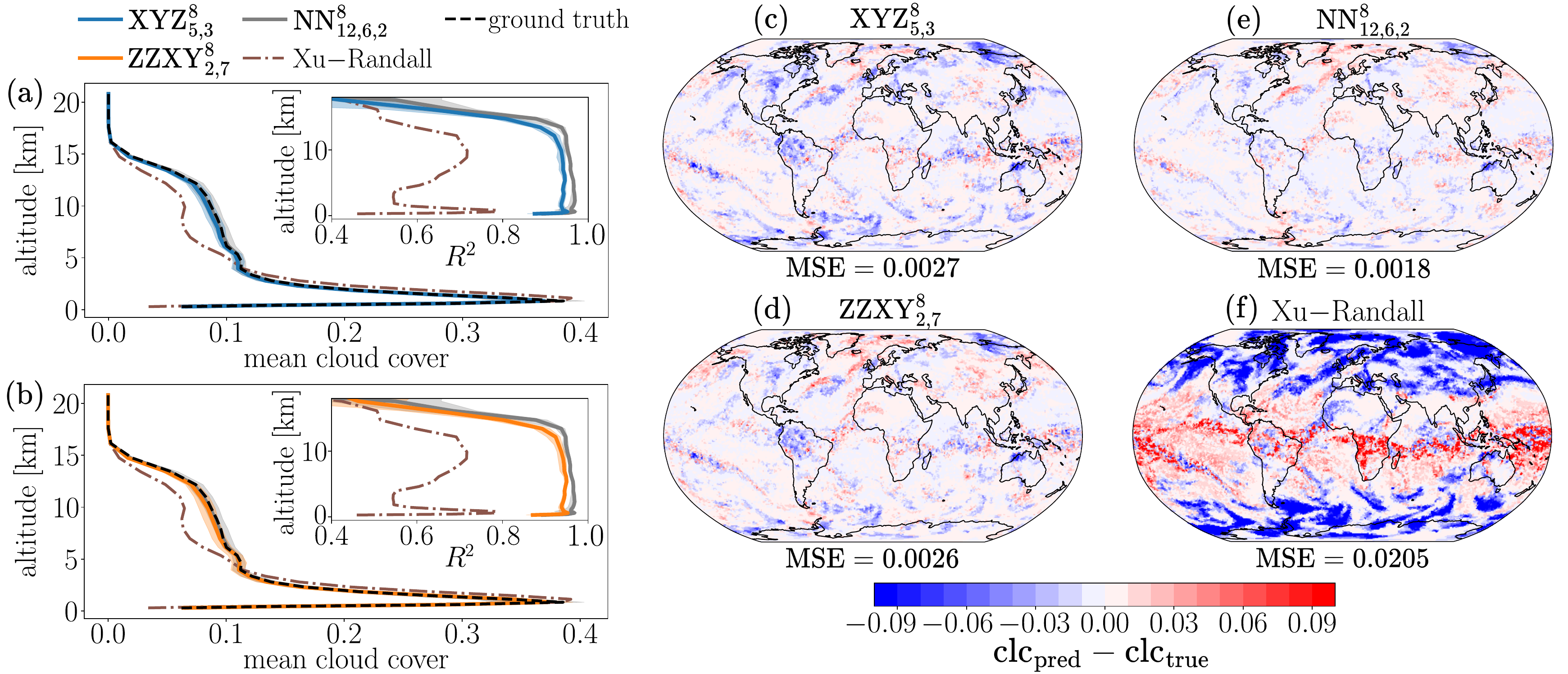}
    \caption{Evaluation of quantum neural networks (QNNs) compared to a classical NN and the Xu-Randall scheme. (a) and (b) show the vertical profiles of the mean cloud cover (i.e., the horizontally averaged cloud cover for each model layer) for the XYZ (blue) and ZZXY (orange) QNNs, respectively. These are compared with the predictions from a classical NN (grey), and from the Xu-Randall scheme (brown). The insets show the corresponding vertically resolved $R^2$ coefficients. The shaded areas correspond to the spread (minimum to maximum) over an ensemble of $20$ training instances (i.e., training the model on the same data but starting from different random initial parameters). Panels (c), (d), (e) and (f) show global maps of the difference between the predicted and the true vertically averaged cloud cover for a specific simulated day (29 February 2020, 15:00 UTC+00:00), for predictions from the XYZ QNN, ZZXY QNN, the classical NN and the Xu-Randall scheme, respectively. All networks are trained for $200$ epochs with $N_{\mathrm{train}}=2\times 10^5$ training data. Remark: the $\mathrm{MSE}$ and $R^2$ shown here are calculated on the non-transformed output, i.e., after applying the inverse output transform $\mathrm{clc}=g^{-1}(f)$.}
    \label{fig:QNNs_NNs_evaluation_plots} 
\end{figure*}

We start by presenting the results of our QNNs trained and evaluated on the DYAMOND dataset, comparing them with classical methods such as classical feed-forward NNs and a traditionally used cloud cover parameterization scheme. To enable a fair comparison between the representational capability of QNNs and classical NNs, we restrict the architectures of the latter to a number of trainable parameters (approximately) equal to that of our QNNs. The specific details of the QNN and NN architectures are given in Table \ref{table:QNNarchs} (with further details on the classical architectures in Appendix \ref{app:classical_NNs}). For quantum and classical NNs, these design choices are optimized for reaching the best performance for a fixed number of trainable parameters. All compared models are given the same input features and are trained with the same number of training data for the same number of epochs. The influence of the number of training data on the performance of the networks and the analysis of the training dynamics is analyzed in the next sections, while here we focus on the evaluation of the predictions of the trained quantum and classical networks. 

The quantum and classical network architectures presented in this section use the eight input features $\{q_{\mathrm{v}}, q_{\mathrm{c}}, q_{\mathrm{i}}, T, p, z_{\mathrm{g}}, h_{\mathrm{w}}, \phi\}$ (corresponding to the $\mathrm{XXY}^8_{5,3}$, $\mathrm{ZZXY}^8_{2,7}$, and $\mathrm{NN}^8_{12,6,2}$ models in Table \ref{table:QNNarchs}).
All networks are trained for $200$ epochs on $N_{\mathrm{train}}=2\times 10^5$ training data from the DYAMOND dataset. After training, we evaluate the networks on a testing dataset $\mathcal{D}_{\mathrm{test}}$, also extracted from the DYAMOND dataset. As for the training set, the testing samples are extracted at random locations in space and time (with care taken in not making them overlap with the training samples). Besides the MSE defined in Eq.~(\ref{eq:MSEdef}), another metric we use to evaluate the performance of our networks is the $R^2$ coefficient of determination defined as:
\begin{equation}
    R^2=1-\frac{\mathrm{MSE}_{\mathcal{D}_{\mathrm{test}}}}{\mathrm{Var}_{\mathcal{D}_{\mathrm{test}}}(\mathrm{clc})}\,\,,
    \label{eq:R2def}
\end{equation}
where $\mathrm{Var}_{\mathcal{D}_{\mathrm{test}}}(\mathrm{clc})$ denotes the variance of the true cloud cover value over the testing dataset. Further evaluation metrics for our networks are discussed in Appendix \ref{app:other_metrics_eval}. For a better assessment of the performance of our XYZ and ZZXY QNNs, we also compare their predictions to a simplified version of the parameterization scheme developed by Xu and Randall \citep{Xu1996,Wang2023}, which is a semi-empirical scheme and defined by:
\begin{equation}
    \mathrm{clc}_{\mathrm{XR}}=\mathrm{min}\big\{1,\,\mathrm{RH}(q_{\mathrm{v}},p,T)^{\beta}(1-\mathrm{e}^{-\alpha(q_{\mathrm{c}}+q_{\mathrm{i}})})\big\} \,\,,
    \label{eq:XuRandall}
\end{equation}
where $\alpha=4.034\times 10^4$ and $\beta=0.9942$ are parameters whose values we optimized (via standard MSE minimization) to reach the best performance over our training dataset. In the above equation, $\mathrm{RH}$ denotes the relative humidity which is calculated as:
\begin{equation}
    \mathrm{RH}(q_{\mathrm{v}},p,T)=\frac{p_{\mathrm{v}}(q_{\mathrm{v}},p)}{p_{\mathrm{s}}(T)}=\frac{p}{p_{\mathrm{s}}(T)}\,\frac{q_{\mathrm{v}}}{0.622+0.378\,q_{\mathrm{v}}} \,,
    \label{eq:RH}
\end{equation}
where $p_{\mathrm{v}}(q_{\mathrm{v}},p)$ is the water vapor pressure and $p_{\mathrm{s}}(T)$ is the saturation vapor pressure (calculated according to \cite{Murray1967}).

The QNN and NN predictions and their comparison with the Xu-Randall scheme are shown in Fig.~\ref{fig:QNNs_NNs_evaluation_plots}.  
Looking at the vertical mean cloud cover profiles, we see that the quantum and classical NNs accurately predict the true profile (denoted by the dashed black line), while the Xu-Randall scheme exhibits visible biases throughout the whole vertical extent up to approximately $17$ km, where the amount of condensate is so low that no cloud cover is diagnosed. Furthermore, quantum and classical networks are comparable also in terms of the prediction spread over the $20$ training instances performed (where each instance corresponds to a training run starting from a randomly initialized parameter set). To better address the differences between the accuracy of QNNs and of classical NNs, we consider the vertically resolved $R^2$ coefficient in the insets. There we can see that, while both types of ansatz achieve a good prediction performance, the classical NN has a slightly better performance with a higher $R^2$ value of approximately $0.01$ throughout the whole vertical extent. The drop in $R^2$ value for all architectures for altitudes above $15$ km is due to the fact that at such high altitudes there is very low probability of observing clouds, meaning that the variance of cloud cover over the testing set at those altitudes is close to zero. Similar conclusions can be drawn the the global bias maps shown in panels (c), (d), (e) and (f) of Fig.~\ref{fig:QNNs_NNs_evaluation_plots}.
As before, we see that the Xu-Randall scheme has the strongest prediction biases, whereas XYZ, ZZXY and NNs all achieve similar performances. In particular, we observe that all these networks show similar spatial distributions of the biases, with NNs achieving slightly lower MSE compared to the quantum architectures. We do not have a conclusive explanation for the slightly better performance of classical NNs over our QNNs for our task. However, we find it plausible that the NNs used here as comparison (which are the best performing ones on our task --- see Appendix \ref{app:classical_NNs} for details on them) may be slightly better suited for our task, in terms of the functional properties they can access (i.e., a moderate inductive bias). In future work, one could attempt to change the type of data encoding in the QNNs, e.g., along the lines of \cite{williams2023}, to potentially yield better performance. Overall, our QNNs show very similar performance to the classical networks with both achieving a high prediction accuracy and moderate cloud cover biases, thus demonstrating that QML can yield suitable models to predict patterns in climate data. In the next sections, we compare our QNNs with classical NNs on further aspects beyond the standard performance metrics.

\subsection{Generalization in quantum and classical NNs} \label{subsec:noiseless_generalization}

\begin{figure}
    \centering
    \includegraphics[width=0.65\linewidth]{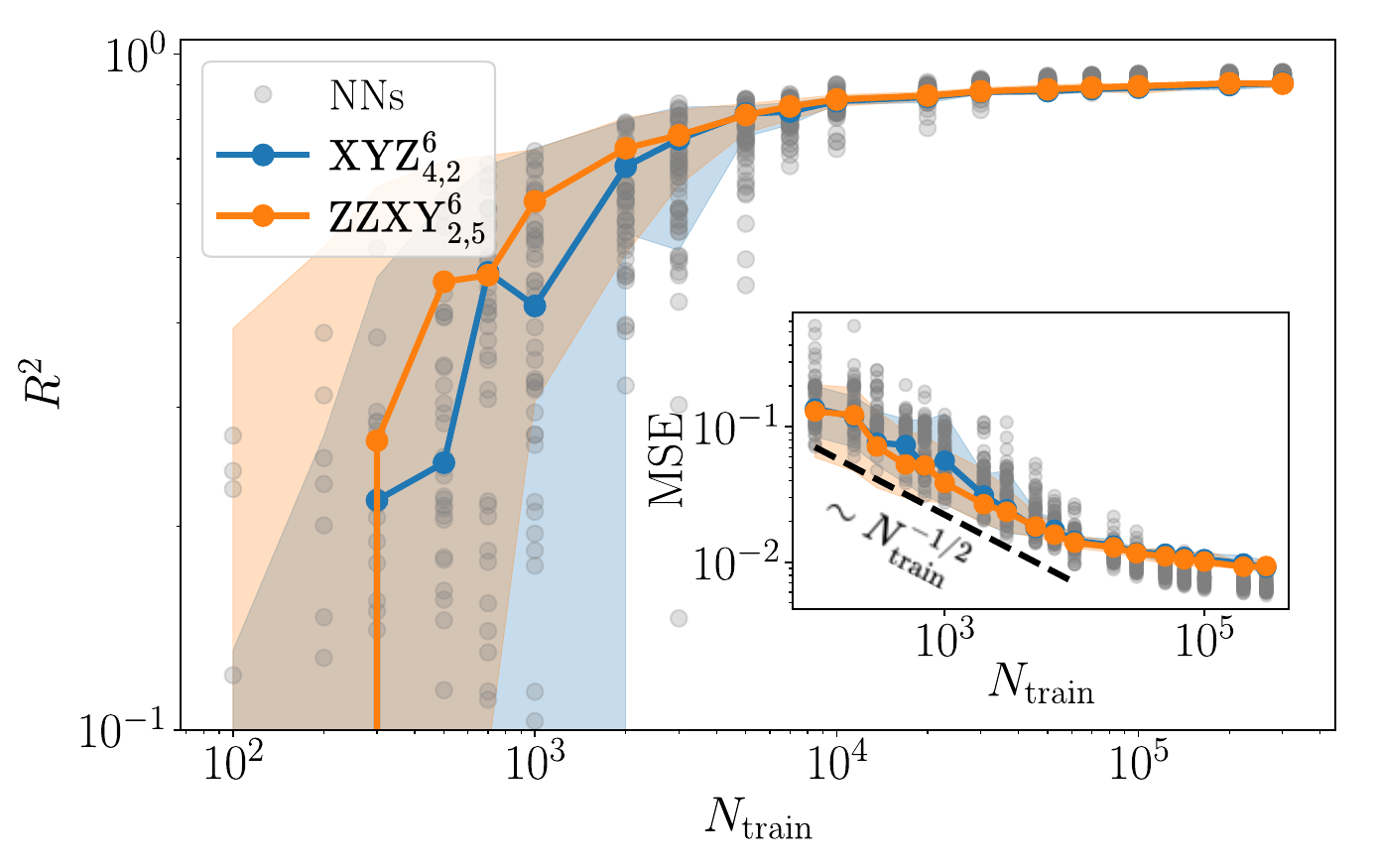}
    \caption{Scaling of network performance with size of the training set $N_{\mathrm{train}}$. The grey points correspond to the tested classical NNs (the six architectures with six input features discussed in Appendix \ref{app:classical_NNs}), with one point corresponding to a specific training instance for the given $N_{\mathrm{train}}$, starting from randomly initialized parameters. The blue and the orange data correspond to the XYZ and the ZZXY QNNs, respectively. We plot $10$ training instances per $N_{\mathrm{train}}$ value and architecture. For the QNNs, the dots indicate the mean over the training instances for a given $N_{\mathrm{train}}$, while the shading denotes the spread (minimum to maximum). The main panel shows the $R^2$ score, the inset the $\mathrm{MSE}$, both calculated over a test set of size $10^5$. 
    Remark: the $\mathrm{MSE}$ and $R^2$ shown here are calculated on the non-transformed output, i.e., after applying the inverse output transform $\mathrm{clc}=g^{-1}(f)$.}
    \label{fig:QNNs_vs_NNs_trainsetsize_scaling} 
\end{figure}

In this section, we address the in-distribution generalization ability of our QNNs and compare it to that of classical NNs. To this end, for both types of networks, we compute the scaling of the performance metrics ($R^2$ and $\mathrm{MSE}$) with the size $N_{\mathrm{train}}$ of the training dataset, where the metrics are computed on a testing dataset $\mathcal{D}_{\mathrm{test}}$ of size $2\times 10^5$ following the same distribution as the training set. Here, we use slightly smaller networks compared to the previous section, to reduce the training runtime of the QNNs for gathering sufficient statistics for our training instances. Specifically, we use the architectures reported in Table \ref{table:QNNarchs} with six input features.
In the comparison, both $R^2$ and the $\mathrm{MSE}$  (Fig.~\ref{fig:QNNs_vs_NNs_trainsetsize_scaling}) show a scaling with $N_{\mathrm{train}}$ that is very similar for the classical and quantum networks. The degradation of the prediction performance happens approximately at the same value of $N_{\mathrm{train}}$, and furthermore no significant difference between our XYZ and ZZXY models can be seen. The $\mathrm{MSE}$ loss on the testing dataset follows approximately a $1/\sqrt{N_{\mathrm{train}}}$ scaling for $N_{\mathrm{train}}\geq D$ until saturation due to model deficiency starts to occur, which is consistent with the scaling laws reported in the works on generalization in QML \citep{Banchi2021,Caro2022a}. Classical and quantum NNs follow approximately the same scaling with the same pre-factors, and, consistently with the observations in the previous section, we see that for large values of $N_{\mathrm{train}}$ the classical NNs achieve a slightly better prediction performance.

\subsection{Trainability from the perspective of information geometry} \label{subsec:noiseless_FIM}

\begin{figure}
    \centering
    \includegraphics[width=0.65\linewidth]{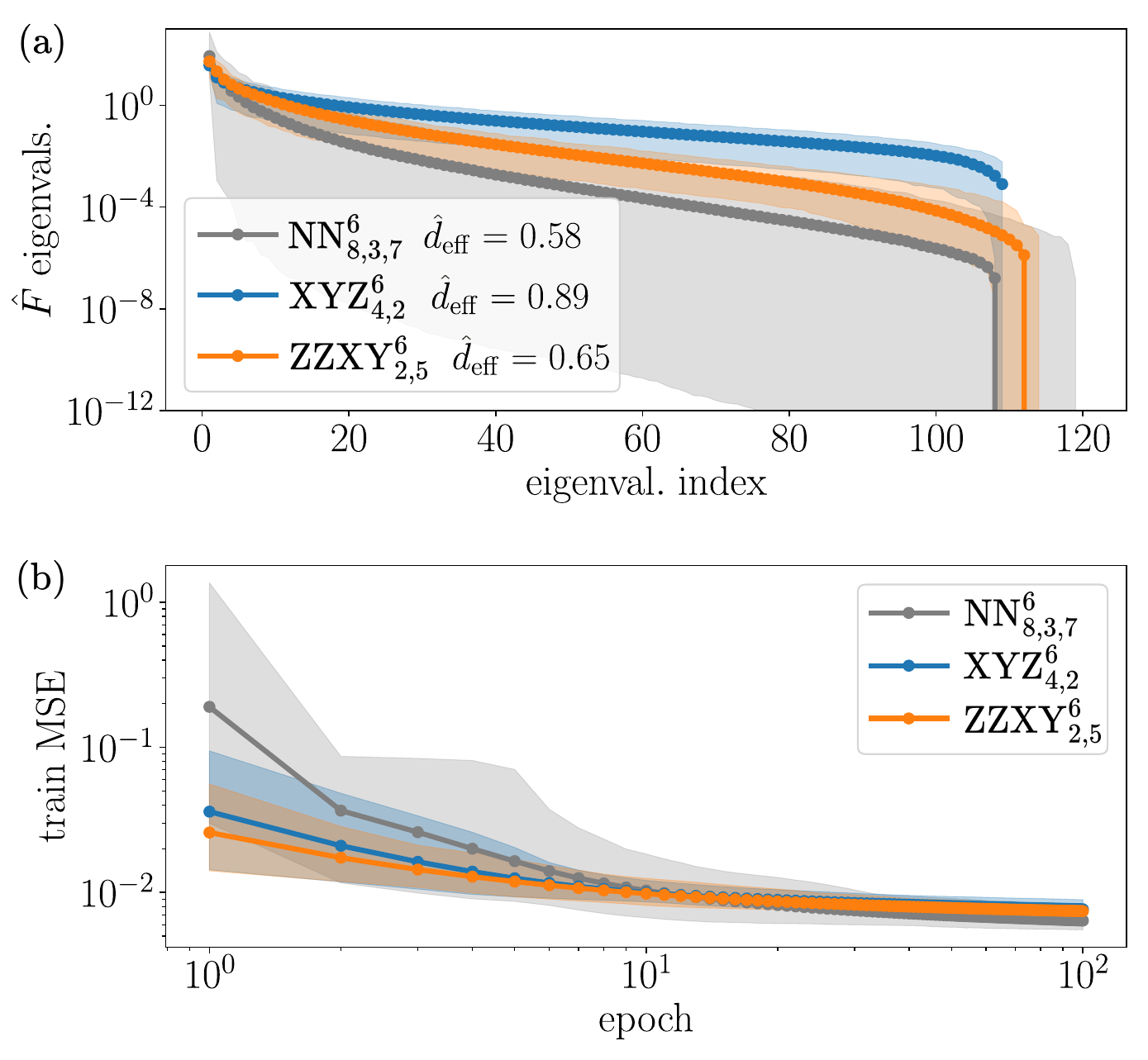}
    \caption{(a) Spectra of normalized FIM for classical NN (grey), and the XYZ (blue) and ZZXY (orange) QNNs. The dots correspond to the mean value over the $200$ random parameter initializations, while the shading denotes the spread (minimum to maximum). To estimate the FIM we use $9\times 10^6$ input data. The normalized effective dimension $\hat{d}_{\mathrm{eff}}$ is also reported in the legend. (b) Training curves for $200$ training instances for classical NN (grey), XYZ (blue) and ZZXY (orange) QNNs. 
    Here, $N_{\mathrm{train}}=10^5$.}
    \label{fig:QNNs_vs_NNs_FIM_and_trainability_plots} 
\end{figure}

In this section, we analyze the training dynamics of our QNNs and NNs, and try to establish a connection with the geometrical properties of their optimization landscape, along the lines of \cite{Abbas2021}. The authors of \cite{Abbas2021} use tools from information geometry, in particular the Fisher information matrix (FIM), setting them in relation with the trainability of a model, and introduce a quantity called ``effective dimension'', which they use to prove generalization bounds. They conduct numerical experiments comparing classical and quantum NNs, and empirically show that QNNs have a better behaved (i.e., more evenly spread) FIM spectrum and a higher effective dimension, which in their experiments results in a faster and more stable training compared to classical networks. Motivated by these empirical observations, here we conduct a similar experiment. Specifically, for each quantum and classical network tested, we draw an ensemble of random parameters configurations (here consisting of $200$ samples), and use it for constructing an ensemble of training instances on the one hand, and to calculate the FIM and the models effective dimension on the other hand. Before presenting the results, we briefly recap the definition and meaning of the FIM and effective dimension.

The FIM describes the geometrical properties of the parameter space of a parameterized model \citep{Amari1985}. It is a central tool in the field of information geometry, as it defines a metric in the model space, i.e., the space of functions a parameterized model can represent \citep{Amari1985,Amari1997,Pascanu2014}. While the FIM is typically defined for probabilistic models, its definition can be extended to the regression models studied here (see \cite{Pennington2018,Karakida2020} and Appendix \ref{app:FIM_details} for details). In this case, the elements of the FIM are computed as follows
\begin{equation}
\begin{split}
    F_{j,k}(\boldsymbol{\theta})&=\mathbb{E}_{\boldsymbol{x}}\bigg[\frac{\partial f_{\boldsymbol{\theta}}(\boldsymbol{x})}{\partial\theta_j}\frac{\partial f_{\boldsymbol{\theta}}(\boldsymbol{x})}{\partial\theta_k}\bigg]\\
    &\approx\frac{1}{|\mathcal{D}|}\sum_{\boldsymbol{x}\in\mathcal{D}}\frac{\partial f_{\boldsymbol{\theta}}(\boldsymbol{x})}{\partial\theta_j}\frac{\partial f_{\boldsymbol{\theta}}(\boldsymbol{x})}{\partial\theta_k} \,\,,
\end{split}
\label{eq:FIM}
\end{equation}
with $\mathbb{E}_{\boldsymbol{x}}$ denoting the expected value over the input distribution, which is empirically approximated by averaging over the (training) dataset $\mathcal{D}$. The FIM $F$ is a $D\times D$ positive semi-definite matrix. Giving rise to a metric in model space, the FIM encodes information on which directions in the parameter space lead to appreciable changes in the outputs of the model, and which directions are instead less influential or redundant. This information is reflected in the spectrum of the FIM: an evenly spread and non-zero spectrum indicates that all parameters are almost equally contributing to independent model changes, whereas the presence of small or zero eigenvalues means that the corresponding parameters (or linear combinations thereof) are irrelevant. Based on this observation, the FIM spectrum becomes a useful indication of the capacity of a model to effectively explore its parameter space, `making use' of all its degrees of freedom. This capacity, according to \cite{Abbas2021}, may also be beneficial during training. The (normalized) effective dimension introduced in \cite{Abbas2021} is constructed to capture this capacity, and is defined as
\begin{equation}
    \hat{d}_{\mathrm{eff}}=\frac{2\,\mathrm{log}\bigg(\frac{1}{V_{\mathrm{\Theta}}}\int_{\mathrm{\Theta}}\sqrt{\mathrm{det}\Big(I_D+c_{N_{\mathrm{data}}}\hat{F}(\boldsymbol{\theta})\Big)}\mathrm{d}\boldsymbol{\theta}\bigg)}{D\,\mathrm{log}\,c_{N_{\mathrm{data}}}} \,\,,
    \label{eq:norm_eff_dim}
\end{equation}
where $c_{N_{\mathrm{data}}}=\frac{N_{\mathrm{data}}}{2\pi\,\mathrm{log}\,N_{\mathrm{data}}}$ with $N_{\mathrm{data}}\equiv|\mathcal{D}|$ being the number of input data samples, $D$ the number of parameters, and $\hat{F}(\boldsymbol{\theta})$, the normalized FIM, defined as
\begin{equation}
    \hat{F}(\boldsymbol{\theta})=\frac{D}{\frac{1}{V_{\mathrm{\Theta}}}\int_{\mathrm{\Theta}}\mathrm{tr}\big(F(\boldsymbol{\theta})\big)\mathrm{d}\boldsymbol{\theta}}\,F(\boldsymbol{\theta}) \,\,,
\end{equation}
with $\frac{1}{V_{\mathrm{\Theta}}}\int_{\mathrm{\Theta}}\mathrm{tr}\big(F(\boldsymbol{\theta})\big)\mathrm{d}\boldsymbol{\theta}\approx\frac{1}{M}\sum_{m=1}^{M}\mathrm{tr}\big(F(\boldsymbol{\theta}_m)\big)$. The $\hat{d}_{\mathrm{eff}}$ is normalized to the dimensionality of the parameter space $D$, hence being bounded in $[0,1]$. Furthermore, it is computed from averages over the parameter space, hence depending solely on the architecture choices and the input distribution.

Equipped with these definitions, we can now present the results of our experiments. 
From the spectra of the normalized FIM shown in Fig.~\ref{fig:QNNs_vs_NNs_FIM_and_trainability_plots}(a) we can make two observations. First, on average the QNNs have larger FIM eigenvalues and a flatter spectrum, and second, the spread over the $200$ parameter samples is larger for NNs compared to the QNNs. These observations are consistent with those of \cite{Abbas2021}, and are reflected in the higher value of $\hat{d}_{\mathrm{eff}}$ for the QNNs, as reported in the legend. 

We now move on to investigate whether these different geometrical properties result in practical differences in the training dynamics of our networks. This is shown in panel (b) of Fig.~\ref{fig:QNNs_vs_NNs_FIM_and_trainability_plots}. Here, we observe that both quantum and classical architectures train to approximately the same value of the loss function, with the classical NN achieving slightly lower loss. The behavior of the training loss, in particular the speed at which the training has approximately converged, is very similar when comparing classical and quantum networks, unlike the observations in \cite{Abbas2021}. Therefore, for our test case we cannot pinpoint a strong relationship between the geometrical properties described before and the effectiveness of training (similar observations have been also raised in a recent work \cite{Mingard2024}). 
We refer the reader to Appendix \ref{app:FIM_details} for a further analysis of the relation between training dynamics and FIM, which also includes tests on other NNs and QNNs.
We conclude this section by mentioning that the trainability aspects investigated here concern only the training dynamics of QNNs for a fixed number of qubits, and that we are not addressing the problem of barren plateaus in this work, expected to occur for general QNN architectures with large numbers of qubits \citep{McClean2018,Thanasilp2023,Larocca2024}.

\section{Influence of shot noise} \label{sec:results_noisy}

In a real quantum device, the expectation values $\langle\hat{\sigma}_n^z\rangle_{\boldsymbol{\vartheta},\boldsymbol{\varphi}}(\boldsymbol{x})$ used in our QNNs would need to be estimated from a finite number of repeated measurements, also called shots, on the output quantum state. This introduces statistical fluctuations, i.e., shot noise, in the estimates for $\langle\hat{\sigma}_n^z\rangle_{\boldsymbol{\vartheta},\boldsymbol{\varphi}}(\boldsymbol{x})$. In this section we analyze the effects of shot noise on the training and performance of our QNNs, and their dependence on the number $n_{\mathrm{shots}}$ of measurement shots. Furthermore, we present an application of the variance regularization technique introduced in \cite{Kreplin2024} to our use case, and show that it can help to reduce the effects of shot noise in the training and subsequent inference stage of our QNN models.

\subsection{Training and inference with shot noise} \label{subsec:noisy_training_and_eval}

\begin{figure}
    \centering
    \includegraphics[width=0.65\linewidth]{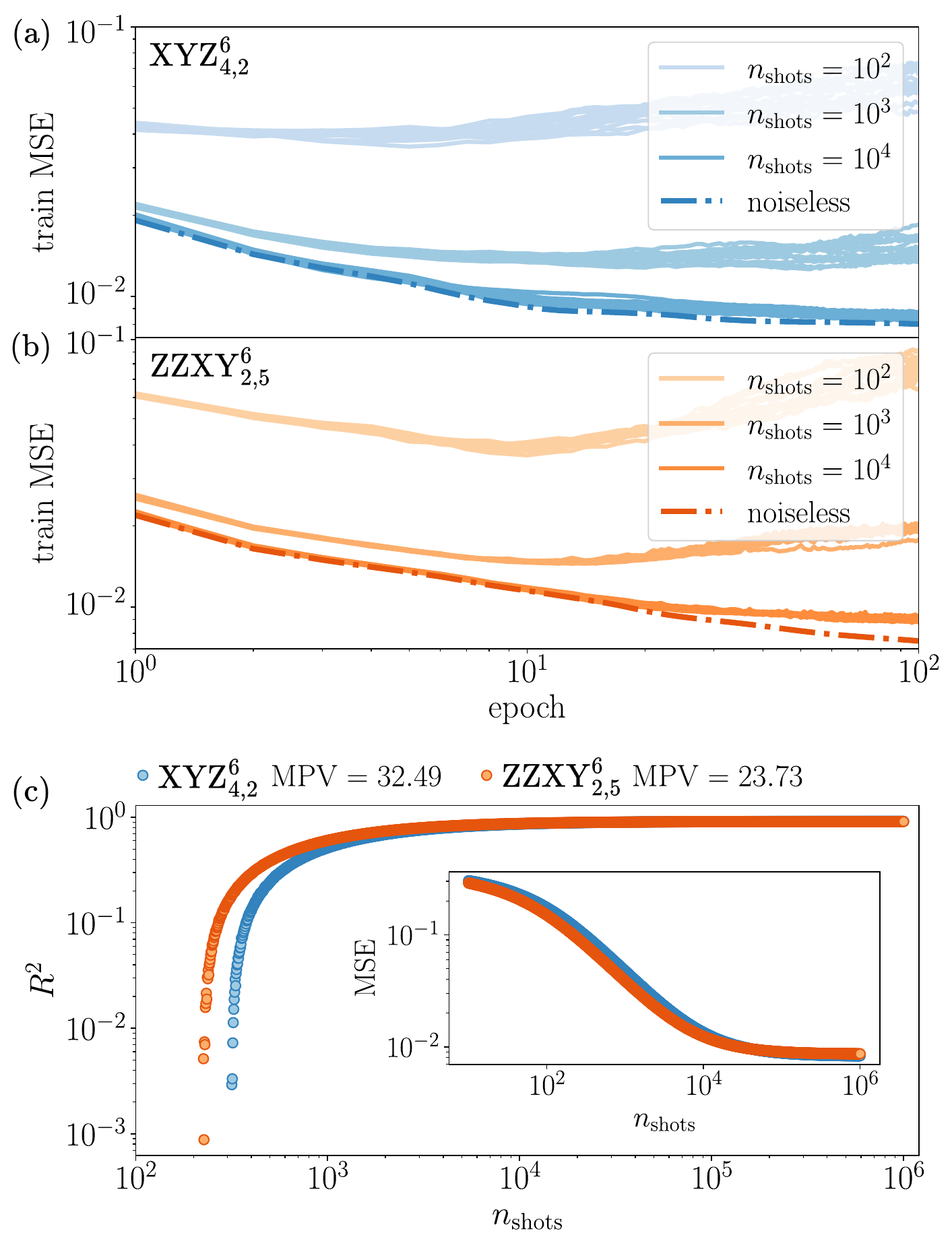}
    \caption{(a) and (b) Training curves for XYZ (blue) and ZZXY (orange) QNNs, respectively, for different values of $n_{\mathrm{shots}}$ used in the evaluation of predictions and gradients. The different solid lines correspond to different training instances performed starting from the same initial parameter configuration, but with different shot noise realizations. The dashed line corresponds to the noiseless limit, i.e., predictions and gradients evaluated to numerical precision. Here, $N_{\mathrm{train}}=10^5$. (c) Test set performance of XYZ (blue) and ZZXY (orange) QNNs trained in the noiseless regime, against $n_{\mathrm{shots}}$ used at inference, for a test set of size $2\times 10^5$. 
    Remark: The $\mathrm{MSE}$ and $R^2$ shown here are calculated on the non-transformed output, i.e., after applying the inverse output transform $\mathrm{clc}=g^{-1}(f)$.}
    \label{fig:QNNs_noisy_training_inference_novarreg} 
\end{figure}

We now analyze the influence of shot noise on the training dynamics and on the prediction performance of our QNNs. To this end, it is instructive to recall how shot noise influences the predictions of our QNNs. The QNNs' predictions $f_{\boldsymbol{\theta}}(\boldsymbol{x})$ are calculated from a weighted average of the expectation values $\langle\hat{\sigma}_n^z\rangle_{\boldsymbol{\vartheta},\boldsymbol{\varphi}}$ which, in a realistic quantum device, would be estimated from a finite number $n_{\mathrm{shots}}$ of measured bit-strings as
\begin{equation*}
    \mathrm{Est}_{n_{\mathrm{shots}}}[\langle\hat{\sigma}_n^z\rangle_{\boldsymbol{\vartheta},\boldsymbol{\varphi}}]=\frac{1}{n_{\mathrm{shots}}}\sum_{s=1}^{n_{\mathrm{shots}}}b_n^{(s)} \,\,,
\end{equation*}
with $b_n^{(s)}\in\{-1,+1\}$ denoting the $n$-th `bit' of the $s$-th measured bit-string. Let us call $z_n\equiv\mathrm{Est}_{n_{\mathrm{shots}}}[\langle\hat{\sigma}_n^z\rangle_{\boldsymbol{\vartheta},\boldsymbol{\varphi}}]$ (we drop here the dependence on $\boldsymbol{x}$, $\boldsymbol{\vartheta}$ and $\boldsymbol{\varphi}$ for notational convenience). The $z_n$ are random variables with covariance given by
\begin{equation*}
    \mathrm{Cov}[z_m,z_n]=\frac{\langle\hat{\sigma}_m^z\hat{\sigma}_n^z\rangle-\langle\hat{\sigma}_m^z\rangle\langle\hat{\sigma}_n^z\rangle}{n_{\mathrm{shots}}} \,\,.
\end{equation*}
Using this, we can calculate the variance of the final prediction as
\begin{equation}
    \mathrm{Var}[f_{\boldsymbol{\theta}}(\boldsymbol{x})]=\sum_{m,n=1}^N w_m\,w_n\,\mathrm{Cov}[z_m,z_n] \,\,,
\end{equation}
with $w_n$ being the $n$-th trainable weight in $\boldsymbol{w}$ (see Eq.~\eqref{eq:final_QNN_pred}). Every QNN evaluation for calculating both predictions and gradients (with the parameter-shift rule) is then affected by statistical fluctuations which in general negatively impact the training and prediction performance of the model.

We start by analyzing the training dynamics with a finite $n_{\mathrm{shots}}$ for the XYZ and ZZXY networks. 
The results are shown in panels (a) and (b) of Fig.~\ref{fig:QNNs_noisy_training_inference_novarreg}, where we plot the training $\mathrm{MSE}$ during the training epochs for different $n_{\mathrm{shots}}$ (fixed throughout the training) for both the XYZ and ZZXY architectures, respectively. The different curves with the same color correspond to different training instances obtained by starting from the same initial parameter set, but with different shot noise realizations. For both architectures it is visible that for $n_{\mathrm{shots}}$ sufficiently large (i.e., on the order of $10^4$) the training is successful, closely following the noiseless training curve. For smaller values of $n_{\mathrm{shots}}$, the $\mathrm{MSE}$ is minimized in the initial iterations before the training becomes unstable, which happens when the gradients' magnitude becomes comparable to the noise in the gradients' evaluation. These results demonstrate that there is a regime where our QNNs train stably in the presence of shot noise, albeit it being relatively costly in the number of necessary circuit evaluations. We discuss in the next section a method targeted towards minimizing these costs.

Second, we analyze the performance of our QNNs on a testing dataset when varying $n_{\mathrm{shots}}$. Specifically, we use the optimal network parameters configurations for both QNNs obtained after training in the noiseless regime, and test how a finite $n_{\mathrm{shots}}$ (in particular smaller than $10^4$) affects their prediction performance. A useful metric we compute to understand the (average) impact of shot noise on the test set is the mean prediction variance (MPV), which is defined on a dataset $\mathcal{D}$ as
\begin{equation}
    \mathrm{MPV}_{\mathcal{D}}(\boldsymbol{\theta})=\frac{1}{|\mathcal{D}|}\sum_{\boldsymbol{x}_i\in\mathcal{D}}\mathrm{Var}[f_{\boldsymbol{\theta}}(\boldsymbol{x}_i)] \,\,.
    \label{eq:mean_pred_var}
\end{equation}
The results are shown in panel (c) of Fig.~\ref{fig:QNNs_noisy_training_inference_novarreg}, where we plot the $R^2$ and the $\mathrm{MSE}$ for different $n_{\mathrm{shots}}$ for the XYZ and ZZXY architectures. From these results it is clear that the quality of the predictions rapidly degrades for $n_{\mathrm{shots}}<10^4$ while it is stable for larger values. The slight difference between the two architectures can be attributed to the difference in the value of the $\mathrm{MPV}$, which is indicated in the plot. Specifically, the ZZXY architecture has a slightly lower $\mathrm{MPV}$, which results in a lower number of shots needed to achieve a given accuracy. The $\mathrm{MPV}$, capturing this difference, becomes therefore a crucial ingredient for the method presented in the next section.

\subsection{Variance regularization} \label{subsec:variance_reg}

\begin{figure*}
    \centering
    \includegraphics[width=1.05\linewidth]{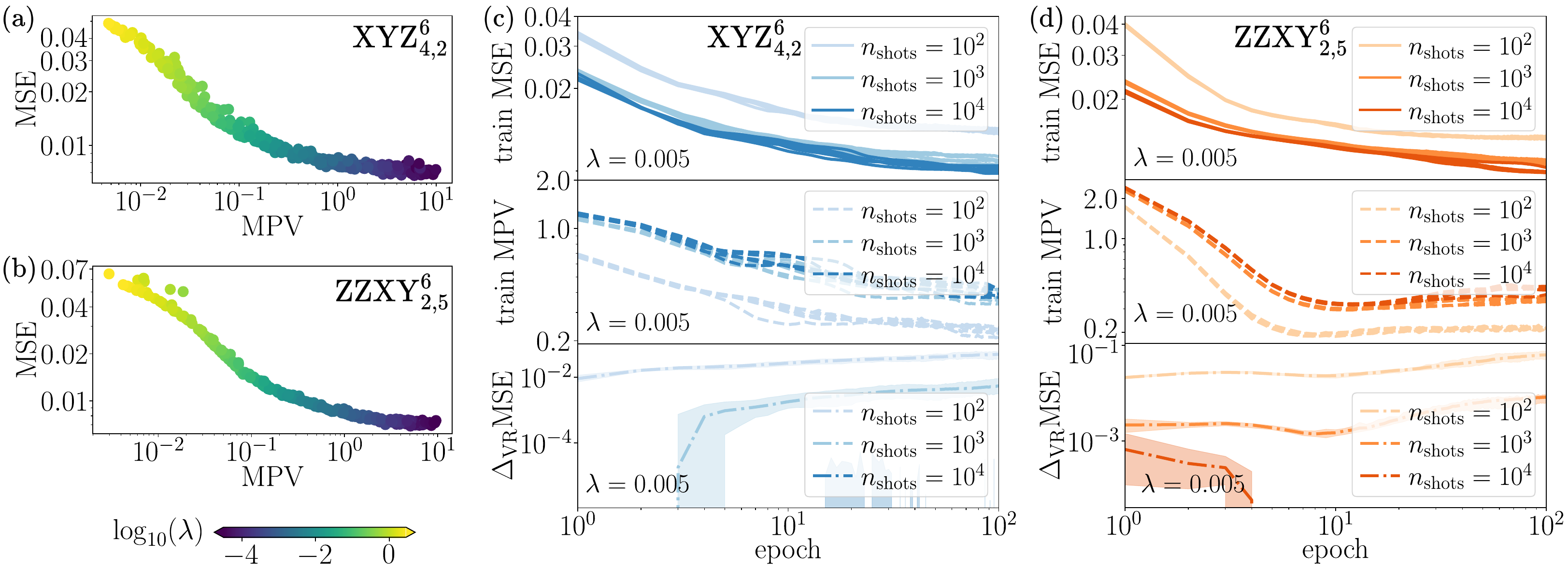}
    \caption{Variance regularization results. (a) and (b) Post-training mean squared error ($\mathrm{MSE}$) and mean prediction variance ($\mathrm{MPV}$) on testing set for XYZ and ZZXY QNNs, respectively. The color scale refers to the value of $\lambda$, the training is performed in the noiseless regime, with $2\times10^5$ training data for $150$ epochs, and $2\times10^5$ testing data points are used. (c) and (d) Noisy training curves using variance regularization with $\lambda=0.005$ for XYZ (c) and ZZXY (d) QNNs, respectively. The upper row (solid lines) shows the mean squared error ($\mathrm{MSE}$), the middle row (dashed lines) shows the mean prediction variance ($\mathrm{MPV}$). The different color shades correspond to different $n_{\mathrm{shots}}$ used, while the six different curves with the same colors correspond to six different training instances (from the same initial parameters, but different shot noise realizations). The lower row (dash-dotted line) shows the difference between the $\mathrm{MSE}$ without variance regularization and the $\mathrm{MSE}$ with variance regularization ($\Delta_{\mathrm{VR}}\mathrm{MSE}\equiv\mathrm{MSE}_{\mathrm{no\,var.\,reg.}}-\mathrm{MSE}_{\mathrm{var.\,reg.}}$). The shading correspond to the spread over the training instances.
    }
    \label{fig:variance_reg} 
\end{figure*}

In this section, we discuss the variance regularization technique introduced in \cite{Kreplin2024} and apply it to our QNNs for cloud cover. Variance regularization is a technique aimed at reducing the effects of measurement shot noise in the training and inference stage of a QML model. This is achieved by adding to the loss function (here the $\mathrm{MSE}$) a term which is proportional to the variance of the output of the model, in order to simultaneously minimize both terms (if possible) and therefore obtain a model requiring fewer shots to be evaluated, both during and after training \citep{Kreplin2024}. In our context, the loss function that we minimize in the training reads as
\begin{equation}
    \mathcal{L}_{\mathcal{D}}(\boldsymbol{\theta};\lambda)=\mathrm{MSE}_{\mathcal{D}}(\boldsymbol{\theta})+\lambda\,\mathrm{MPV}_{\mathcal{D}}(\boldsymbol{\theta}) \,\,,
\end{equation}
where $\mathrm{MPV}_{\mathcal{D}}(\boldsymbol{\theta})$ is defined in Eq.~\eqref{eq:mean_pred_var} and $\lambda$ is a regularization parameter. In general, the evaluation of $\mathrm{MPV}_{\mathcal{D}}(\boldsymbol{\theta})$ requires measurements in addition to those performed for evaluating the model output. In our case however, we can estimate both $\mathrm{MSE}$ and $\mathrm{MPV}$ from the same set of measurement shots, since $f_{\boldsymbol{\theta}}(\boldsymbol{x})$ is constructed only from $\hat{\sigma}^z$ expectation values. Calculating the gradients of $\mathcal{L}_{\mathcal{D}}(\boldsymbol{\theta};\lambda)$ can be done with the same parameter-shift rule of Eq.~\eqref{eq:param_shift_rule}, therefore this changing of the loss does not result in additional training overhead. 

We test this technique on our two QNN architectures in the same setting as in the previous section. The regularization parameter $\lambda$ is chosen to be constant during training, although we remark here that \cite{Kreplin2024} also proposes a dynamical regularization approach which may further improve the training. The results are shown in Fig.~\ref{fig:variance_reg}. Training the QNNs with variance regularization in the noiseless regime (Fig.~\ref{fig:variance_reg} (a) and (b)) shows that there exist values of $\lambda$ for which the $\mathrm{MPV}$ can be reduced of more than one order of magnitude while still keeping the $\mathrm{MSE}$ low (i.e., below $10^{-2}$). In other words, these results show that there are regions of the QNNs' parameter space where both the prediction error and the prediction variance can be low, and that variance regularization can effectively target them provided one chooses a suitable value for $\lambda$. 

With these positive results, we move on to applying variance regularization with a finite $n_{\mathrm{shots}}$, to assess whether a finite $\lambda$ can help in stabilizing the training even when $n_{\mathrm{shots}}<10^4$ (Fig.~\ref{fig:variance_reg} (c) and (d)). From both panels it is evident that a value of $\lambda=0.005$ is already sufficient to stabilize the training even for a relatively low $n_{\mathrm{shots}}=100$. For a better comparison with the case of no variance regularization, we also show the difference with respect to the training $\mathrm{MSE}$ in the previous section (Fig.~\ref{fig:QNNs_noisy_training_inference_novarreg}(a)), in the lower row of panels (c) and (d). For $n_{\mathrm{shots}}<10^4$, the difference $\Delta_{\mathrm{VR}}\mathrm{MSE}\equiv\mathrm{MSE}_{\mathrm{no\,var.\,reg.}}-\mathrm{MSE}_{\mathrm{var.\,reg.}}$ remains positive throughout the training, indicating a better training performance when using a finite $\lambda$. For a large number of shots $n_{\mathrm{shots}}\geq10^4$, the difference is very small or negative, which indicates that in this regime adding the $\mathrm{MPV}$ to the loss contrasts the $\mathrm{MSE}$ minimization. For further minimizing the $\mathrm{MSE}$ one could continue the training potentially gradually decreasing $\lambda$, as mentioned above and done in \cite{Kreplin2024}. Overall, the results presented here show that variance regularization is an effective approach for achieving a stable training of a QNN even with a moderate number of shots.

\section{Discussion and outlook} \label{sec:discussion}
In this paper we explore the potential of QNNs as parameterization schemes in climate models, focusing on the specific case of cloud cover parameterizations. With the goal of predicting cloud cover from coarse-grained state variables, we compare different QNN architectures with classical NNs of similar size (i.e., similar number of parameters) on several aspects: prediction accuracy, generalization ability and trainability. Overall, the investigated QNNs show a similar performance (at least in the noiseless setting) to classical NNs in all these aspects, which is a promising indication of the applicability of these models to learn patterns in climate data. Our work opens up several directions for further investigating the applicability of QML in developing parameterizations for climate models. We outline those in the following, while also discussing criticalities and potential limitations associated to them.

A first extension to our work would be to study the performance and learning behavior of the proposed architectures for problems of larger size, which would require moving away from the cell-based approach we took here, by considering also neighboring model cells (as done in \cite{Grundner2022} in the context of classical NNs). Furthermore, while we have performed a thorough analysis of possible QNN ansatzes, we are aware that our architecture search is not exhaustive. Notable interesting examples to investigate in future works could include quantum convolutional neural networks \citep{Cong2019} or QNNs with measurement feedforwards \citep{FossFeig2023,Sahay2024,Chan2024,Iqbal2024}. While increasing the size of the problem, and thus the number of qubits, may introduce trainability issues such as barren plateaus with increasing number of qubits \citep{McClean2018,Thanasilp2023,Larocca2024}, we note that there exist architectures that are immune to such problems \citep{Pesah2021}, and parameter initialization strategies to mitigate them \citep{Grant2019,Friedrich2022,Zhang2022,Gelman2024}.

Our construction could be also tested on parameterization schemes other than cloud cover. Changing the parameterization scheme would likely result in a more complex learning task, which may also increase the chance of observing a separation between the quantum and classical ML algorithms for the considered dataset. In fact, it is to a large extent unknown what types of classical datasets are better suited to quantum learning models than to a classical ones, and extensive tests on several architectures and datasets as done here are very valuable to pinpoint features that may help answering this question.

Going beyond regression, another interesting extension would be to re-frame the parameterization learning problem in a probabilistic setting. Specifically, one could build quantum generative models \citep{Amin2018,DallaireDemers2018,Coyle2020,Gao2022} for learning the full probability distribution of the process considered (here cloud cover). Switching to generative modeling, a task that is to some extent more natural to quantum computing, may increase the chance of observing a separation between the quantum and classical models for the given problem \citep{Du2020}.

Finally, another direction concerns the online implementation of the proposed QNNs, i.e., their coupling to the dynamical core of the climate model solving the Navier-Stokes equations. While QNNs consisting of a small number of qubits can be numerically simulated and thus in principle be readily used within a climate model, a larger number of qubits would require the coupling of a quantum device to the dynamical core. Given that parameterization schemes need to be run for every cell of the model at every time-step, such a coupling could become impractical in terms of computation runtime. To address this limitation, a possibility would be to build classical surrogates \citep{Schreiber2023,Landman2022,Sweke2023,Jerbi2024} of the trained QNNs, to be then used online. In this way, the quantum device would be used only in the development stage of the parameterization, and our work lays a solid basis for developing such a workflow. Another possible direction are quantum-inspired methods such as tensor networks \citep{Orus2019}, which already find applications as (Q)ML models \citep{Stoudenmire2016,Efthymiou2019,Haghshenas2022,Ran2023,Rieser2023}, in classical data compression and loading \citep{Dilip2022,Jumade2023,Jobst2023}, and generative modeling \citep{Han2018,Merbis2023,Liu2023}.

\backmatter

\bmhead{Data and code availability}
The code and the data used for this work can be made available from the authors upon reasonable request.

\bmhead{Acknowledgements}
This project was made possible by the DLR Quantum Computing Initiative and the Federal Ministry for Economic Affairs and Climate Action; qci.dlr.de/projects/klim-qml. A.G.~and V.E.~were funded by the European Research Council (ERC) Synergy Grant “Understanding and Modelling the Earth System with Machine Learning (USMILE)” under the Horizon 2020 research and innovation programme (Grant agreement No.~855187).
V.E.~was additionally supported by the Deutsche Forschungsgemeinschaft (DFG, German Research Foundation) through the Gottfried Wilhelm Leibniz Prize awarded to Veronika Eyring (Reference No.~EY 22/2-1).
This work used resources of the Deutsches Klimarechenzentrum (DKRZ) granted by its Scientific Steering Committee (WLA) under project ID bd1179.

\begin{appendices}

\section{Other QNN architectures tested} \label{app:other_QNNs}
In this appendix we provide the details of the other QNN architectures tested in our work but not shown in the main text. The CNOT-PBC architecture contains as entangling gates CNOT gates arranged in a chain ring pattern. The encoding layer for this architecture is $\hat{S}_{\mathrm{CNOT\mathit{-}PBC}}(\boldsymbol{x})=\hat{R}_{x}(\boldsymbol{x})$. The encoding blocks for the CNOT-PBC read as
\begin{equation*}
\begin{split}
    \hat{V}_{\mathrm{CNOT\mathit{-}PBC}}(\boldsymbol{\vartheta})=\,&\hat{R}_{z}(\boldsymbol{\vartheta}_{(N+1)\to 2N})\times\\
    &\hat{R}_{y}(\boldsymbol{\vartheta}_{1\to N})\,\widehat{\mathrm{CNOT}}_{\mathrm{PBC}} \,\,,
\end{split}
\end{equation*}
with $\widehat{\mathrm{CNOT}}_{\mathrm{PBC}}=\widehat{\mathrm{CNOT}}_{N,1}\prod_{n=1}^{N-1}\widehat{\mathrm{CNOT}}_{n,n+1}$, and the variational blocks as
\begin{equation*}
\begin{split}
    \hat{W}_{\mathrm{CNOT\mathit{-}PBC}}(\boldsymbol{\varphi})=\,&\hat{R}_{x}(\boldsymbol{\varphi}_{(2N+1)\to 3N})\,\hat{R}_{z}(\boldsymbol{\varphi}_{(N+1)\to 2N})\times \\
    &\hat{R}_{y}(\boldsymbol{\varphi}_{1\to N})\,\widehat{\mathrm{CNOT}}_{\mathrm{PBC}} \,\,.
\end{split}
\end{equation*}
The CNOT-NN architecture contains as entangling gates CNOT gates acting between neighboring qubits in a chain. The encoding layer for this architecture is $\hat{S}_{\mathrm{CNOT\mathit{-}NN}}(\boldsymbol{x})=\hat{R}_{x}(\boldsymbol{x})$. The encoding blocks for the CNOT-NN read as
\begin{equation*}
    \hat{V}_{\mathrm{CNOT\mathit{-}NN}}(\boldsymbol{\vartheta})=\hat{R}_{z}(\boldsymbol{\vartheta}_{(N+1)\to 2N})\,\hat{R}_{y}(\boldsymbol{\vartheta}_{1\to N})\,\widehat{\mathrm{CNOT}}_{\mathrm{NN}} \,\,,
\end{equation*}
with $\widehat{\mathrm{CNOT}}_{\mathrm{NN}}=\prod_{n=1}^{N-1}\widehat{\mathrm{CNOT}}_{n,n+1}$, and the variational blocks as
\begin{equation*}
\begin{split}
    \hat{W}_{\mathrm{CNOT\mathit{-}NN}}(\boldsymbol{\varphi})=\,&\hat{R}_{x}(\boldsymbol{\varphi}_{(2N+1)\to 3N})\,\hat{R}_{z}(\boldsymbol{\varphi}_{(N+1)\to 2N})\times \\
    &\hat{R}_{y}(\boldsymbol{\varphi}_{1\to N})\,\widehat{\mathrm{CNOT}}_{\mathrm{NN}} \,\,.
\end{split}
\end{equation*}
The IONS architecture contains as entangling operation one that is naturally implemented in trapped ions quantum simulators, generated by a long-range Hamiltonian of the form $\sum_{n=1}^{N-1}\sum_{m<n}\frac{\hat{\sigma}^x_n\,\hat{\sigma}^x_m}{m-n}$, coming from the laser coupling of the ions internal states to the center-of-mass vibrational mode of the ion chain. The IONS architecture takes as input the state $\prod_{n=1}^N\hat{\mathrm{H}}_n\ket{0}$ with $\hat{\mathrm{H}}_n$ being the Hadamard gate on qubit $n$. The encoding layer for this architecture is $\hat{S}_{\mathrm{IONS}}(\boldsymbol{x})=\hat{R}_{z}(\boldsymbol{x})$.
The encoding blocks for the IONS architecture read as
\begin{equation*}
    \hat{V}_{\mathrm{IONS}}(\boldsymbol{\vartheta})=\hat{R}_{y}(\boldsymbol{\vartheta}_{2\to(N+1)})\,\hat{U}_{\mathrm{IONS}}(\vartheta_1) \,\,,
\end{equation*}
with $\hat{U}_{\mathrm{IONS}}(\vartheta)=\exp(-\frac{\mathrm{i\vartheta}}{2}\sum_{n<m}\frac{\hat{\sigma}^x_n\,\hat{\sigma}^x_m}{m-n})$, and the variational blocks
\begin{equation*}
\begin{split}
    \hat{W}_{\mathrm{IONS}}(\boldsymbol{\varphi})=\,&\hat{R}_{y}(\boldsymbol{\varphi}_{(2N+2)\to(3N+1)})\,\hat{U}_{\mathrm{IONS}}(\varphi_{(2N+1)})\,\times \\
    &\hat{R}_{z}(\boldsymbol{\varphi}_{(N+1)\to 2N})\,\hat{R}_{x}(\boldsymbol{\varphi}_{1\to N}) \,\,.
\end{split}
\end{equation*}
Additionally, we also experimented with augmenting the CNOT-PBC and CNOT-NN architectures with higher-order angle encoding layers of the form $\hat{S}_{\mathrm{HONE}}(\boldsymbol{x})=\hat{R}_{x}(\boldsymbol{x}^{[2]})\,\hat{R}_{x}(\boldsymbol{x})$, with $\boldsymbol{x}^{[2]}$ being a vector with $(N-1)$ components which are computed from $\boldsymbol{x}$ as $x^{[2]}_m=\frac{x_m\,x_{m+1}}{2\pi}$.

\section{Details on input and output transformations} \label{app:input_transf}

\begin{figure*}
    \centering
    \includegraphics[width=1.05\linewidth]{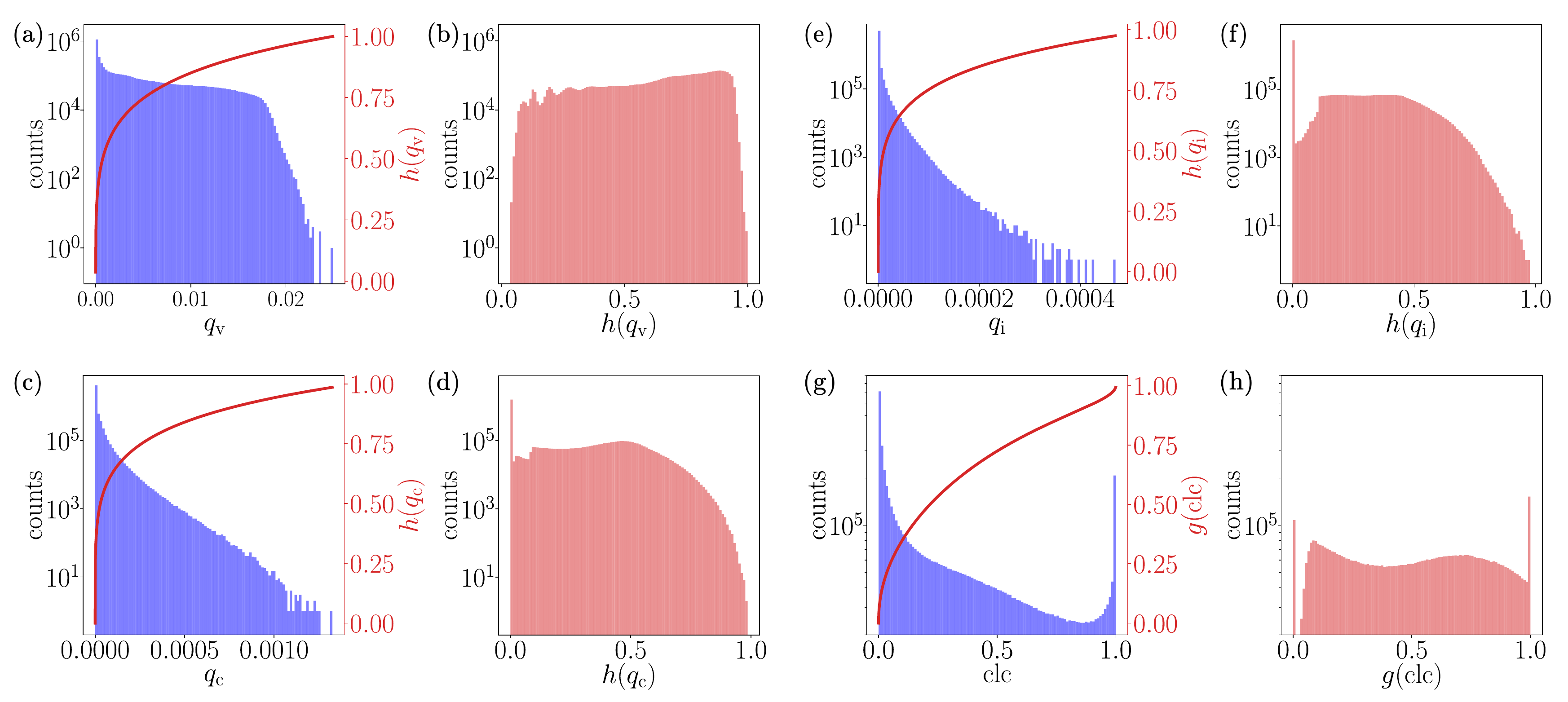}
    \caption{Input and output transformation functions and histograms. (a) Histogram of specific humidity $q_{\mathrm{v}}$ in the training dataset (blue) and transformation function $h(x)$ (red). (b) Histogram of the transformed $h(q_{\mathrm{v}})$. (c) Histogram of cloud water $q_{\mathrm{c}}$ in the training dataset (blue) and transformation function $h(x)$ (red). (d) Histogram of the transformed $h(q_{\mathrm{c}})$. (e) Histogram of cloud ice $q_{\mathrm{i}}$ in the training dataset (blue) and transformation function $h(x)$ (red). (f) Histogram of the transformed $h(q_{\mathrm{i}})$. (g) Histogram of cloud cover $\mathrm{clc}$ in the training dataset (blue) and transformation function $g(x)$ (red). (h) Histogram of the transformed $g(\mathrm{clc})$.}
    \label{fig:in_out_transf_app} 
\end{figure*}

In this appendix we discuss the input and output transformations that we applied to the DYAMOND data before feeding them to our classical and quantum models. 
For the input features, the idea behind the transformation is to make the feature distribution more uniform within a specified interval, and to still retain the input feature variability in the tails, which can be associated with physical scenarios we are interested in capturing. The transformation function we designed in this case reads as
\begin{equation*}
h(x)=\frac{\log\bigg[1+(\mathrm{e}-1)\Big(\frac{x}{x_{\mathrm{high}}}\Big)^b\bigg]-h_0(b,x_{\mathrm{low}},x_{\mathrm{high}})}{1-h_0(b,x_{\mathrm{low}},x_{\mathrm{high}})} \,\,,
\end{equation*}
where $h_0(b,x_{\mathrm{low}},x_{\mathrm{high}})=\log\bigg[1+(\mathrm{e}-1)\Big(\frac{x_{\mathrm{low}}}{x_{\mathrm{high}}}\Big)^b\bigg]$ and $x_{\mathrm{low}},x_{\mathrm{high}}$ corresponding to the (approximate) minimum and maximum value of the given feature $x$ estimated on the training dataset. We used the following parameters for the input features:
\begin{itemize}
    \item Specific humidity $q_{\mathrm{v}}$ [kg/kg]: $b=0.25$, $x_{\mathrm{low}}=10^{-7}$, $x_{\mathrm{high}}=0.025$,
    \item Cloud water $q_{\mathrm{c}}$ [kg/kg]: $b=0.25$, $x_{\mathrm{low}}=0$, $x_{\mathrm{high}}=0.00145$,
    \item Cloud ice $q_{\mathrm{i}}$ [kg/kg]: $b=0.25$, $x_{\mathrm{low}}=0$, $x_{\mathrm{high}}=0.00055$,
    \item Horizontal wind $h_{\mathrm{w}}$ [m/s]: $b=0.5$, $x_{\mathrm{low}}=0.0015$, $x_{\mathrm{high}}=115.0$.
\end{itemize}
The remaining input features are transformed using a simple min-max scaling, and all features (including those listed above) are scaled within the interval $[0,\pi]$ (i.e., we multiplied the above $h(x)$ by a factor $\pi$). The transformations for $q_{\mathrm{v}}$, $q_{\mathrm{c}}$ and $q_{\mathrm{i}}$ and the resulting histograms of the transformed values are shown in Fig.~\ref{fig:in_out_transf_app}. In our experiments, the input features are then rescaled to the interval $[0,\pi]$.

Also the output transformation function $g(x)$ is constructed in order to have the training outputs (targets) in a more uniform distribution compared to the original one in the DYAMOND dataset, which in our case improved the performance of both our quantum and classical models. The transformation function is invertible, and reads as
\begin{equation*}
g(x)=\frac{1}{2}+\frac{1}{\pi}\,\arcsin\bigg[2\bigg(\frac{\mathrm{e}^{b\,x^a}-1}{\mathrm{e}^b-1}\bigg)^c-1\bigg]
\end{equation*}
with parameters $a=1.29407913$, $b=-3.20011015$, $c=0.70308237$, which have been chosen in order to have approximate uniformity. The transformation and the resulting transformed outputs histogram are shown in panels (g) and (h) of Fig.~\ref{fig:in_out_transf_app}.

\section{Details on classical NNs tested} \label{app:classical_NNs}
In this Appendix we detail the structure of the classical NNs used as comparison for our QNNs. The NNs presented here are the result of an extensive architecture search with the constraint of keeping the number of parameters approximately equal to that of the QNNs discussed in the main text. We start from the networks taking as inputs the eight input features $\{q_{\mathrm{v}}, q_{\mathrm{c}}, q_{\mathrm{i}}, T, p, z_{\mathrm{g}}, h_{\mathrm{w}}, \phi\}$, and containing a number $D$ of trainable parameters between $200$ and $210$. In the following lists, the number denotes the number of nodes in the given layer, and in parentheses we write the activation function used.
\begin{itemize}
    \item 8 (inputs) $\to$ 10 (tanh) $\to$ 7 (tanh) $\to$ 4 (tanh) $\to$ 1 (linear - output).
    \item 8 (inputs) $\to$ 9 (tanh) $\to$ 4 (tanh) $\to$ 9 (tanh) $\to$ 4 (tanh) $\to$ 1 (linear - output).
    \item 8 (inputs) $\to$ 8 (tanh) $\to$ 8 (tanh) $\to$ 6 (tanh) $\to$ 1 (linear - output).
    \item 8 (inputs) $\to$ 12 (tanh) $\to$ 6 (tanh) $\to$ 2 (tanh) $\to$ 1 (linear - output): best performing, chosen for Fig.~\ref{fig:QNNs_NNs_evaluation_plots}.
    \item 8 (inputs) $\to$ 10 (tanh) $\to$ 10 (tanh) $\to$ 1 (linear - output).
    \item 8 (inputs) $\to$ 8 (tanh) $\to$ 8 (tanh) $\to$ 4 (tanh) $\to$ 4 (tanh) $\to$ 1 (linear - output).
\end{itemize}
For the networks taking as inputs the six input features $\{q_{\mathrm{v}}, q_{\mathrm{c}}, q_{\mathrm{i}}, T, p, h_{\mathrm{w}}\}$, containing a number $D$ of trainable parameters between $109$ and $120$, we used the following layouts.
\begin{itemize}
    \item 6 (inputs) $\to$ 8 (tanh) $\to$ 3 (tanh) $\to$ 7 (tanh) $\to$ 1 (linear - output): best performing, chosen for Fig.~\ref{fig:QNNs_vs_NNs_FIM_and_trainability_plots}.
    \item 6 (inputs) $\to$ 6 (tanh) $\to$ 5 (tanh) $\to$ 5 (tanh) $\to$ 1 (linear - output).
    \item 6 (inputs) $\to$ 7 (tanh) $\to$ 3 (tanh) $\to$ 7 (tanh) $\to$ 2 (tanh) $\to$ 1 (linear - output).
    \item 6 (inputs) $\to$ 8 (tanh) $\to$ 3 (leaky-ReLU) $\to$ 3 (tanh) $\to$ 2 (tanh) $\to$ 2 (leaky-ReLU) $\to$ 2 (tanh) $\to$ 1 (linear - output).
    \item 6 (inputs) $\to$ 10 (tanh) $\to$ 4 (tanh) $\to$ 1 (linear - output).
    \item 6 (inputs) $\to$ 5 (tanh) $\to$ 5 (tanh) $\to$ 4 (tanh) $\to$ 4 (tanh) $\to$ 1 (linear - output).
\end{itemize}
All the architectures in the last list are the ones also used for generating the plots in Fig.~\ref{fig:QNNs_vs_NNs_trainsetsize_scaling} and \ref{fig:QNNs_NNs_training_metrics_app}. For these architectures as well as for the quantum ones, the initial learning rate for the Adam optimizer is set to $0.001$ and the batch size to $100$, which are approximately optimal settings in both classical and quantum cases for our problem.

\begin{figure*}
    \centering
    \includegraphics[width=1.05\linewidth]{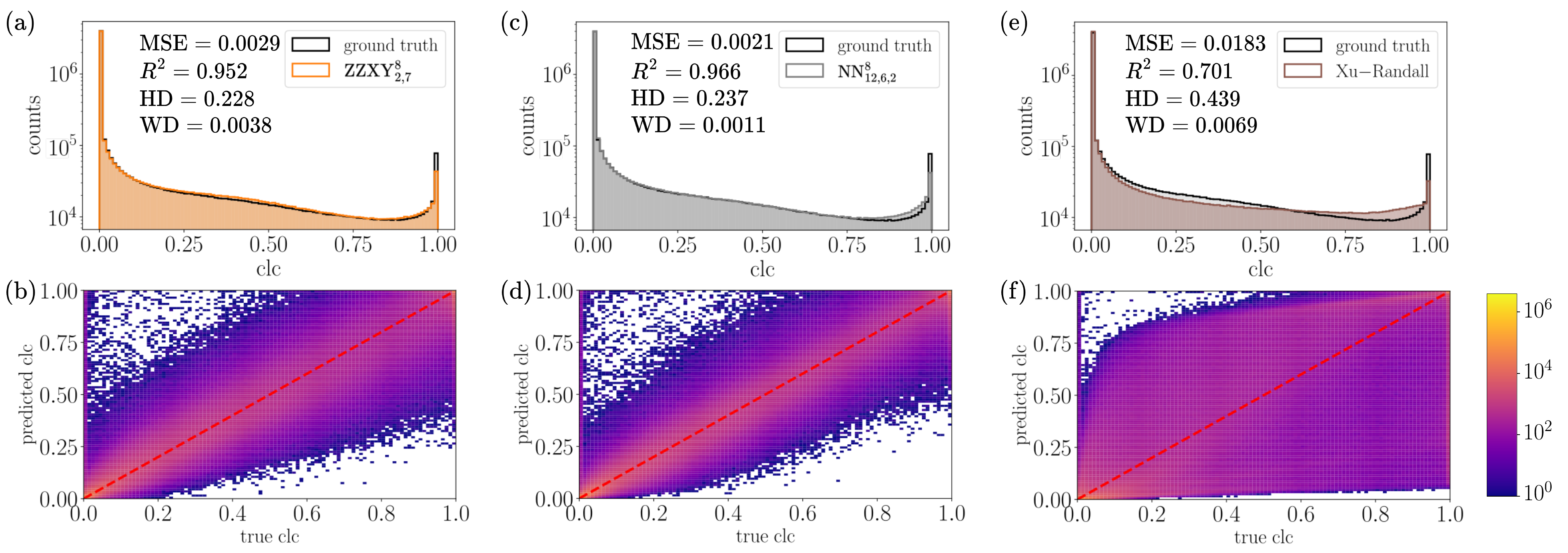}
    \caption{Output distributions. (a) Histogram of predicted (blue) and true (black) $\mathrm{clc}$ from ZZXY QNN. (b) Histogram of true vs.~predicted $\mathrm{clc}$ for ZZXY QNN (logarithmic color scale). (c) Histogram of predicted (orange) and true (black) $\mathrm{clc}$ from classical NN. (d) Histogram of true vs.~predicted $\mathrm{clc}$ for classical NN (logarithmic color scale). (e) Histogram of predicted (brown) and true (black) $\mathrm{clc}$ from Xu-Randall scheme. (f) Histogram of true vs.~predicted $\mathrm{clc}$ for Xu-Randall scheme (logarithmic color scale). The histograms are obtained for a testing set of $6\times 10^6$ data points. We report also the values of the mean squared error (MSE), $R^2$ coefficient, Hellinger distance (HD) and Wasserstein distance (WD) on the same testing set.}
    \label{fig:histograms_app} 
\end{figure*}

\section{Other evaluation metrics} \label{app:other_metrics_eval}

In this appendix we discuss further metrics that can be used to evaluate our quantum and classical networks beyond $\mathrm{MSE}$ and $R^2$. The additional metrics we compute measure the distance between the distributions of the model predictions and the distribution of cloud cover in a testing dataset. Specifically, we calculate the Hellinger distance:
\begin{equation*}
    \mathrm{HD}(P,Q)=\frac{1}{\sqrt{2}}\big\Vert\sqrt{P}-\sqrt{Q}\big\Vert_2 \,\,,
\end{equation*}
where $P$ and $Q$ are the predicted and the true (discretized) probability distributions for cloud cover, and $\Vert\cdot\Vert_2$ denotes the Euclidean norm. Additionally, we calculate the Wasserstein distance
\begin{equation*}
    \mathrm{WD}(P,Q)=\inf_{\Pi\in\Gamma(P,Q)}\mathbb{E}_{(x,y)\sim\Pi}\big(\Vert x-y\Vert\big) \,\,,
\end{equation*}
where $\Gamma(P,Q)$ is the set of all joint probability distributions that have $P$ and $Q$ as marginals (we calculate $\mathrm{WD}$ using the SciPy package in Python).

\begin{figure}
    \centering
    \includegraphics[width=0.65\linewidth]{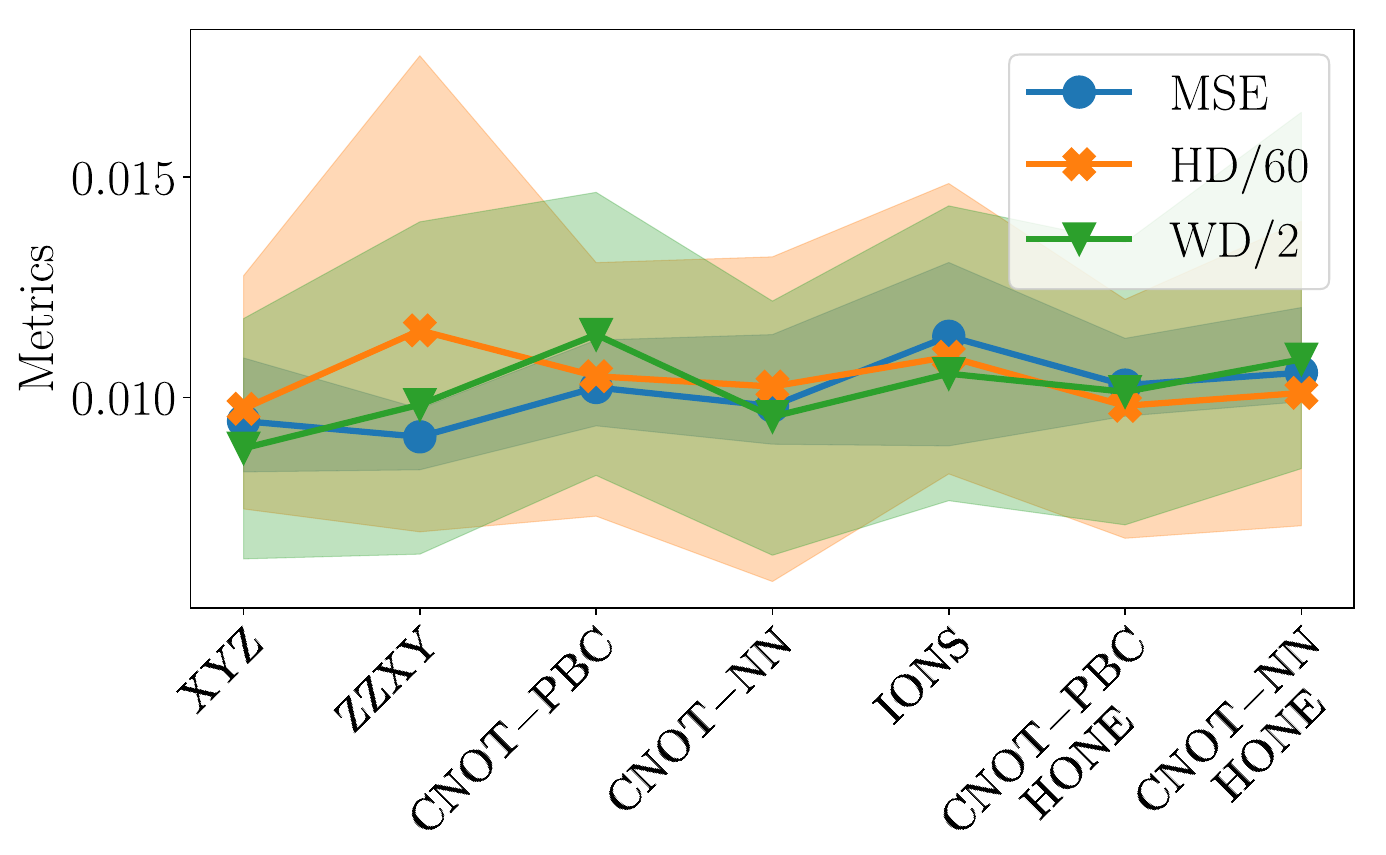}
    \caption{Evaluation metrics for the different QNNs tested: mean squared error (MSE, in blue), Hellinger distance (HD, in orange) and Wasserstein distance (WD, in green). HD and WD have been rescaled in order for all the metrics to have values in the same order of magnitude (scaling factor in the legend). The shaded area corresponds to the spread over $20$ training instances. All QNNs for this plot have six input features $\{q_{\mathrm{v}}, q_{\mathrm{c}}, q_{\mathrm{i}}, T, p, h_{\mathrm{w}}\}$, the number of trainable parameters is between $110$ and $120$, and are trained with $2\times 10^5$ training data for $150$ epochs.}
    \label{fig:metrics_allQNNs_app} 
\end{figure}

In Fig.~\ref{fig:histograms_app} we show the histograms of the outputs from our ZZXY and NN architectures (the XYZ architecture has very similar performance and we do not show it here), compared with the Xu-Randall scheme, for the QNNs with eight input features used in Section \ref{subsec:noiseless_training_and_eval}, and report also the values of the additional distance metrics just discussed. 
The histograms together with the additional metrics further confirm the observations made in the main text, namely, that the quantum and classical NNs perform comparably on the task at hand, with the classical ones being slightly better, while both outperform the Xu-Randall scheme.
In Fig.~\ref{fig:metrics_allQNNs_app} we show these metrics together with the $\mathrm{MSE}$ for all the QNNs tested in this work, including those not shown in the main text but described in Appendix \ref{app:other_QNNs}. Here we can observe that all QNNs tested have comparable performance in the noiseless case, which is why we show only two types of QNNs in the main text.

\section{Details on FIM calculation and other trainability metrics} \label{app:FIM_details}

\begin{figure*}
    \centering
    \includegraphics[width=1.05\linewidth]{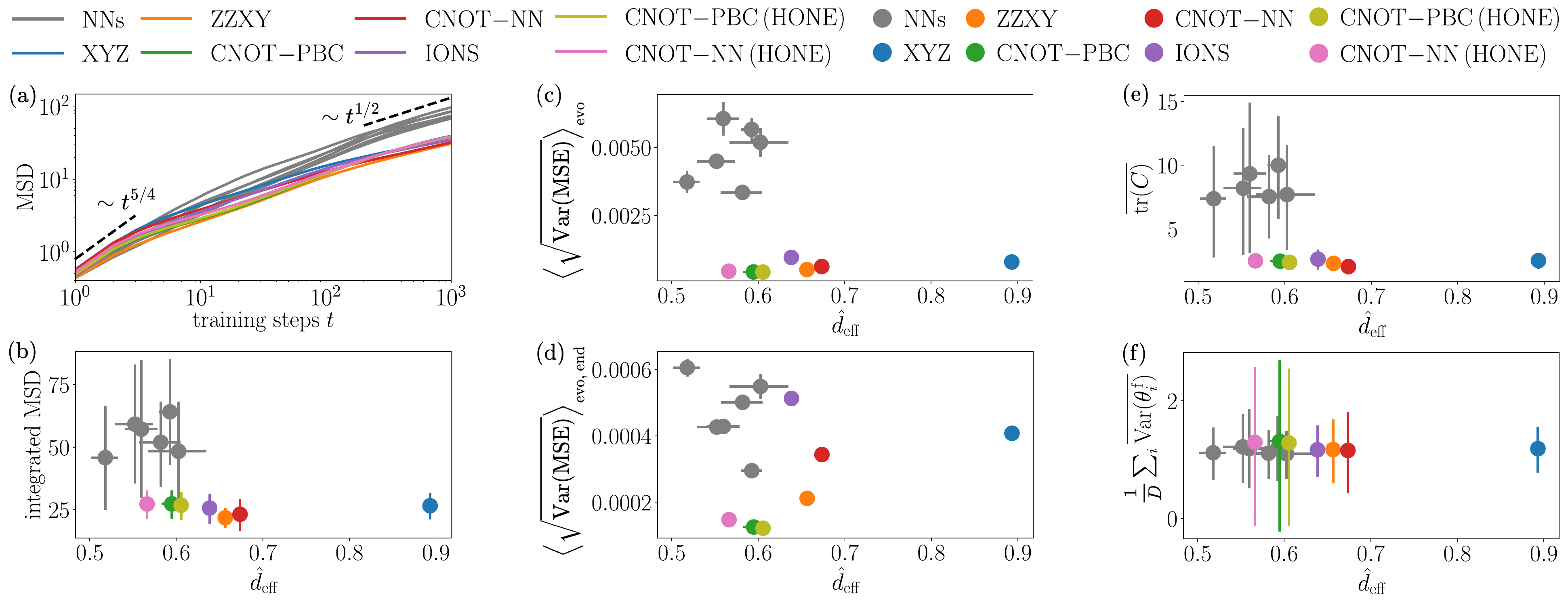}
    \caption{Analysis of training dynamics of classical and quantum networks. (a) Mean squared displacement (MSD) evolution during training. Here each step corresponds the evolution after $100$ batch updates during the Adam optimization (we use $10^5$ training samples, so each epoch consists of $1000$ batch updates with a batch size of $100$). The dashed lines are guides to the eye indicating specific slopes. (b) Integrated MSD (over time) against normalized effective dimension $\hat{d}_{\mathrm{eff}}$. (c) Spread of $\mathrm{MSE}$ loss function averaged over all training epochs plotted against $\hat{d}_{\mathrm{eff}}$. (d) Spread of $\mathrm{MSE}$ loss function averaged over the final 25 training epochs plotted against $\hat{d}_{\mathrm{eff}}$. (e) Trace of parameter correlation matrix $C$ averaged over all $200$ training instances (walkers), plotted against $\hat{d}_{\mathrm{eff}}$. (f) Mean spread of final parameters plotted against $\hat{d}_{\mathrm{eff}}$. The error bars correspond to the estimated statistical uncertainty bootstrapped from the $200$ random parameter realizations. All networks have six input features and the number of trainable parameters is between $109$ and $120$.}
    \label{fig:QNNs_NNs_training_metrics_app} 
\end{figure*}

In this appendix, we give more details on the calculation of the Fisher information matrix (FIM) presented in the main text, and provide a further analysis of the training dynamics of our networks. We start by deriving the formula used for calculating the FIM in our case of regression with MSE loss function. For a statistical model $p(\boldsymbol{x},y;\boldsymbol{\theta})$ the elements of the FIM are defined as \citep{Amari1985,Amari1997,Pascanu2014}
\begin{equation}
F_{j,k}(\boldsymbol{\theta})=\mathbb{E}_{(\boldsymbol{x},y)\sim p}\bigg[\frac{\partial\,\mathrm{log}\,p(\boldsymbol{x},y;\boldsymbol{\theta})}{\partial\theta_j}\frac{\partial\,\mathrm{log}\,p(\boldsymbol{x},y;\boldsymbol{\theta})}{\partial\theta_k}\bigg] \,\,,
\end{equation}
which can be rewritten as 
\begin{equation}
F_{j,k}(\boldsymbol{\theta})=\mathbb{E}_{(\boldsymbol{x},y)\sim p}\bigg[\frac{\partial\,\mathrm{log}\,p_{\boldsymbol{\theta}}(y|\boldsymbol{x})}{\partial\theta_j}\frac{\partial\,\mathrm{log}\,p_{\boldsymbol{\theta}}(y|\boldsymbol{x})}{\partial\theta_k}\bigg] \,\,,
\end{equation}
by noticing that $p(\boldsymbol{x},y;\boldsymbol{\theta})=p(\boldsymbol{x})p_{\boldsymbol{\theta}}(y|\boldsymbol{x})$, with $p_{\boldsymbol{\theta}}(y|\boldsymbol{x})$ being the model output probability conditioned on the input $\boldsymbol{x}$. The quantity $\ell(y,\boldsymbol{x};\boldsymbol{\theta})=-\log p_{\boldsymbol{\theta}}(y|\boldsymbol{x})$ corresponds to the negative log-likelihood, a typical loss function used for statistical models. In our case, with our networks outputting a deterministic value $f_{\boldsymbol{\theta}}(\boldsymbol{x})$ the loss function corresponds to the MSE, which amounts to setting $p_{\boldsymbol{\theta}}(y|\boldsymbol{x})=\mathcal{N}_{f_{\boldsymbol{\theta}}(\boldsymbol{x}),\sigma^2}(y)$ for a fictitious $\sigma$ \citep{Pennington2018,Karakida2020}. Therefore, using
$\partial_{\theta_j}p_{\boldsymbol{\theta}}(y|\boldsymbol{x})=\sigma^{-2}(f_{\boldsymbol{\theta}}(\boldsymbol{x})-y)\,\partial_{\theta_j}f_{\boldsymbol{\theta}}(\boldsymbol{x})$, we can rewrite the FIM as
\begin{equation}
\begin{split}
F_{j,k}(\boldsymbol{\theta})=\mathbb{E}_{\boldsymbol{x}}\bigg[\int&\mathcal{N}_{f_{\boldsymbol{\theta}}(\boldsymbol{x}),\sigma^2}(y)\,(f_{\boldsymbol{\theta}}(\boldsymbol{x})-y)^2\,\sigma^{-4}\,\times\\
&\frac{\partial f_{\boldsymbol{\theta}}(\boldsymbol{x})}{\partial\theta_j}\frac{\partial f_{\boldsymbol{\theta}}(\boldsymbol{x})}{\partial\theta_k}\,\mathrm{d}y\bigg] \,\,.
\end{split}
\end{equation}
Performing the Gaussian integration over $y$, we finally arrive at
\begin{equation}
    F_{j,k}(\boldsymbol{\theta})=\sigma^{-2}\,\mathbb{E}_{\boldsymbol{x}}\bigg[\frac{\partial f_{\boldsymbol{\theta}}(\boldsymbol{x})}{\partial\theta_j}\frac{\partial f_{\boldsymbol{\theta}}(\boldsymbol{x})}{\partial\theta_k}\bigg] \,\,,
\end{equation}
which is proportional to the definition given in Eq.~\eqref{eq:FIM}.

We now move on to discussing further aspects of the training dynamics of our networks beyond the training curves shown in Fig.~\ref{fig:QNNs_vs_NNs_FIM_and_trainability_plots} in the main text. To this end, we not only monitor the value of the $\mathrm{MSE}$ loss during training, but also track the dynamics of the parameters $\boldsymbol{\theta}$, viewing each training experiment starting from a random parameter configuration as a  walker in the parameter space. We denote with $\boldsymbol{\theta}^{(m)}(t)$ ($m=1,...,M$) the parameter configuration of the $m$-th walker at a selected training step $t$. A quantity that is natural to analyze when looking at ensembles of random walkers is the mean squared displacement (MSD), defined as
\begin{equation*}
    \mathrm{MSD}(t)=\overline{\Vert\boldsymbol{\theta}(t)-\boldsymbol{\theta}(0)\Vert^2}\equiv\frac{1}{M}\sum_{m=1}^M \Vert\boldsymbol{\theta}^{(m)}(t)-\boldsymbol{\theta}^{(m)}(0)\Vert^2 \,\,,
\end{equation*}
see Fig.~\ref{fig:QNNs_NNs_training_metrics_app} (a). We observe the same qualitative behavior for both classical and quantum networks, with the initial dynamics being faster (super-diffusive), then progressively slowing down to a sub-diffusive regime when the gradients have significantly decreased in magnitude. Interestingly, it appears that for the QNNs tested here this slow-down of the training dynamics happens faster compared to classical NNs. Attempting to correlate this observation with the geometrical properties of the models captured by the FIM, we consider the integrated MSD, $\frac{1}{T}\sum_{t=1}^T\mathrm{MSD}(t)$ as function of the normalized effective dimension $\hat{d}_{\mathrm{eff}}$ defined in Eq.~\eqref{eq:norm_eff_dim} (Fig.~\ref{fig:QNNs_NNs_training_metrics_app} (b)). While we still observe a difference between classical and quantum architectures, no clear correlation between the the integrated MSD and the value of $\hat{d}_{\mathrm{eff}}$ can be seen. The absence of a clear correlation between the training dynamics and $\hat{d}_{\mathrm{eff}}$ can also be seen in panels (c) and (e). In panel (c) we plot the average of the spread of the $\mathrm{MSE}$ loss during training, which we compute as
\begin{equation*}
    \Big\langle\sqrt{\mathrm{Var}(\mathrm{MSE})}\Big\rangle_{\mathrm{evo}}=\frac{1}{N_{\mathrm{epochs}}}\sum_{j=1}^{N_{\mathrm{epochs}}}\sqrt{\mathrm{Var}(\mathrm{MSE}(j))} \,\,,
\end{equation*}
where $\mathrm{Var}(\mathrm{MSE}(j))$ is the variance of the training $\mathrm{MSE}$ at epoch $j$, calculated over the different walkers. In panel (e) we show the trace of parameter correlation matrix $C$ averaged over all $M$ walkers, as a measure of the cumulative variance of the parameters during their evolution. Specifically, the parameters correlation matrix for the $m$-th walker is calculated as
\begin{equation*}
    C^{(m)} = \frac{1}{T}\,\Theta^{(m)\,\top}\Theta^{(m)} \,\,,
\end{equation*}
where $\Theta^{(m)}$ is a $(T\times D)$ centered data matrix with elements $\Theta^{(m)}_{t,i}=\theta^{(m)}_j(t)-\frac{1}{T}\sum_t\theta^{(m)}_j(t)$. The eigenvectors of $C^{(m)}$ associated to the largest eigenvalues denote the directions along which the parameters have changed the most, and the associated eigenvalues their variance. Hence
\begin{equation*}
    \overline{\mathrm{tr}(C)} \equiv \frac{1}{M}\sum_{m=1}^M\mathrm{tr}(C^{(m)})
\end{equation*}
gives an estimate of the parameters cumulative variance during the evolution. Despite the two quantities sharing similar behavior, from panels (c) and (e) no correlation with $\hat{d}_{\mathrm{eff}}$ can be concluded. The same holds when focusing on only the final part of the training dynamics, as shown in panel (d), as well as when looking at the mean spread of the final parameters $\frac{1}{D}\sum_{i=1}^D\mathrm{Var}(\theta^{\mathrm{f}}_i)$, with $\theta^{\mathrm{f}}_i=\theta_i(T)$ and the variance taken over the walkers ensemble, which is shown in panel (f). Thus, to summarize, in our analysis we see no clear correlation between the geometrical properties captured by the FIM and the training dynamics of our networks.

In particular, we expect the effectiveness of training to be dependent not only on the model ability to effectively explore the parameter space, but also on how well the functions it can represent are suited to the problem at hand (i.e., its inductive bias), an aspect to which the FIM is agnostic. Hence, given the slightly better performance of NNs on our task, their structure may be more favorable for learning cloud cover, which in turn has a positive impact on the training dynamics.

\end{appendices}

\bibliography{QNN4CLC.bib}

\end{document}